\documentclass{article}
\usepackage[onecolumn]{emulateapj,epsfig}

\slugcomment{To appear in ApJ}
\makeatletter

\newenvironment{inlinefigure}{%
\def\@captype{figure}%
\noindent\begin{minipage}{0.999\linewidth}\begin{center}}
{\end{center}\end{minipage}\smallskip}
\makeatother

\newcommand{\HII}{H{\sc ii} }
\newcommand{\HeI}{He{\sc i} }
\newcommand{\HeII}{He{\sc ii} }
\newcommand{\HeIII}{He{\sc iii} }

\newcommand{\HeHII}{{\rm HeH}^+}

\newcommand{\Hthreeplus}{{\rm H}_3^+}

\newcommand{\Hminus}{H$^-$}
\newcommand{\Htwoplus}{{\rm H}_2^+}
\newcommand{\Htwo}{{\rm H}_2}
\newcommand{\HH}{H$_2$ }
\newcommand{\HHp}{H$_2^+$ }

\begin{document}

\title{Early Cosmological HII/HeIII Regions and Their Impact on
Second-Generation Star Formation}

\author{Naoki Yoshida}
\affil{Department of Physics, Nagoya University, Furocho, Nagoya, Aichi 464-8602, Japan}
\author{S. Peng Oh}
\affil{Physics Department, University of California, Santa Barbara, CA 93106}
\author{Tetsu Kitayama}
\affil{Department of Physics, Toho University, Funabashi, Chiba 274-8510, Japan}
\and
\author{Lars Hernquist}
\affil{Harvard-Smithsonian Center for Astrophysics, 60 Garden Street, Cambridge, MA 02138}

\begin{abstract}
We present the results of three-dimensional radiation-hydrodynamics simulations
of the formation and evolution of early H{\sc ii}/\HeIII regions around the 
first stars. Cooling and recollapse of the gas in the relic \HII region is 
also followed in a full cosmological context, until second-generation stars 
are formed. We first carry out ray-tracing simulations of ionizing radiation 
transfer from the first star. Hydrodynamics is directly coupled with 
photo-ionization heating as well as radiative and chemical cooling. 
The photo-ionized hot gas is evacuated out of the host halo at a velocity of 
$\sim$ 30 km/sec. This radiative feedback effect quenches further 
star-formation within the halo for over tens to a hundred million years. 
We show that the thermal and chemical evolution of the photo-ionized gas in the 
relic \HII region is remarkably different from that of a neutral primordial gas. 
Efficient molecular hydrogen production in the recombining gas
enables it to cool to $\sim 100$ K, where fractionation of HD/\HH occurs. 
The gas further cools by HD line cooling down to a few tens Kelvin. 
Interestingly, at high redshifts ($z>10$), the minimum gas temperature is limited 
by that of the cosmic microwave background with $T_{\rm CMB} = 2.728 (1+z)$. 
The gas cloud goes run-away collapse when its mass is
$\sim 40 M_{\odot}$, which is significantly smaller than a typical clump mass of 
$\sim 200-300 M_{\odot}$ for early primordial gas clouds. We argue 
that massive, rather than very massive, primordial stars may form in the relic 
\HII region. Such stars might be responsible for early metal-enrichment of the 
interstellar medium from which recently discovered hyper metal-poor stars were born. 
\end{abstract}


\section{Introduction}
The cosmic dark ages ended when the first sources of light turned on. 
These objects contributed to cosmic reionization, which 
observations suggest began about 
a few hundred million years after the Big Bang (Spergel et al. 2006;
Page et al. 2006), but the exact nature of the sources and how the
process evolved are yet unknown.
In the hierarchical structure formation model based on
Cold Dark Matter (CDM), reionization is characterized by the emergence 
of early \HII regions around individual sources (stars, galaxies or quasars),
followed by percolation of the ionized regions 
(Gnedin \& Ostriker 1997; Miralda-Escud\'e et al. 2000;
Ciardi, Ferrara \& White 2003; Sokasian et al. 2003, 2004; 
Furlanetto et al. 2004a, 2006; Kuhlen \& Madau 2005).
The shape and extent of the early \HII regions determine the 
global topology of the distribution of neutral and ionized gas at 
different epochs (e.g. McQuinn et al. 2005, 2006; Zahn et al. 2005, 2006),
which can be probed by future ground-based radio observations;
turned around, the observations will provide rich
information on the nature of the first sources of light
(Madau, Meiksin \& Rees 1997; Zaldarriaga et al. 2004;
Furlanetto et al. 2004b; Mellema et al. 2006).

The majority of recent theoretical models indicate that a 
dominant contribution 
to the ultra-violet photons that reionized the Universe came from high-redshift, 
low-mass galaxies. These models, however, rely on the crucial assumption that 
star-formation in such low-mass galaxies is as efficient as in the observed local 
Universe. While such an assumption can reproduce the inferred Thomson optical 
depth to electron scattering and the neutral fraction of the 
intergalactic medium at $z\sim 6$, it is necessary to examine whether or not the 
gas within these high redshift galaxies can cool and condense rapidly to enable 
efficient star-formation.

The standard CDM model predicts that the first cosmological 
objects form very early in low-mass halos in which the primordial gas 
condenses via molecular hydrogen cooling
(Couchman \& Rees 1986; Tegmark et al. 1997; Yoshida et al. 2003).
This mass scale is smaller than the often assumed characteristic mass 
for galaxy formation, for which more efficient atomic cooling is thought
to be vital. Therefore, in the hierarchical model, feedback effects from the first 
generation of stars are expected to play a key role in setting the scene, 
{\it i.e. the initial conditions}, for (proto-)galaxy formation.
Intriguingly, recent theoretical studies of the formation of the first stars
(Abel, Bryan \& Norman 2002; Omukai \& Palla 2003; Yoshida et al. 2006)
consistently indicate that these objects were rather massive. It is
then expected that radiation from the first stars would have a considerable impact 
on the thermal and chemical evolution of the surrounding gas cloud and even the 
intergalactic medium. In light of this, Kitayama, Yoshida, Susa \& Umemura (2004, 
hereafter KYSU) performed radiation-hydrodynamics calculations
of early HII regions and determined the critical mass
of complete ionization of halo gas by the central source.
KYSU showed that nearly all the gas is evacuated out of the host
halo by radiation and thermal pressure from
photo-ionization by the central massive star(s). Whalen et al. (2004)
arrived at the same conclusion for a specific case. The long-term evolution
of the relic \HII regions may critically determine the efficiency of 
star-formation in the same place, which is of considerable cosmological 
importance. The evolution of early \HII regions are intrinsically
coupled with the evolution of the surrounding gas and that of the host dark 
matter halos, and thus are indeed a very complicated problem. 

There is another interesting question about the formation of the so-called 
second-generation stars.  Unlike the somewhat simple and `clean' initial 
conditions for the first stars, second generation stars are likely born 
from a gas that has been disturbed by earlier feedback effects.
%newly added by naoki
Among the most important 
of these effects is photo-ionization by the first stars.
An ionized gas cools faster than it recombines, and thus many
free electrons are left over (Shapiro \& Kang 1987; Susa et al. 1998).
When the temperature drops below $\sim 8000$K, atomic hydrogen Lyman-$\alpha$
cooling becomes inefficient but then hydrogen molecules are rapidly
formed using the abundant free electrons as a catalyst. The gas cools
by \HH line cooling to $\sim 100$ K. In a primordial gas, HD molecules 
act as an efficient coolant at such low temperatures, enabling the
gas to cool down to a few tens Kelvin (Flower et al. 2000).
MacLow \& Shull (1986) and Uehara \& Inutsuka (2000) consider 
the evolution of shock-heated gas and argue that cooling by HD molecules is 
important under a broad range of conditions. The latter authors further 
suggested that HD cooling may lead to the formation of primordial brown dwarfs. 
Nakamura \& Umemura (2002) studied the cooling of filamentary
gas using an extended 14 primordial species chemistry that included 
HD and its ions. They identified a critical molecular hydrogen fraction
of $\sim 10^{-3}$; a gas with abundant hydrogen molecules can cool below 150K,
and then formation, and hence cooling by HD molecules 
becomes important in the final run-away collapse phase.
Interestingly, this critical fraction is close to the universal 
asymptotic molecular hydrogen fraction calculated by Susa et al. (1998)
and Oh \& Haiman (2002) for a cooling gas in halos with $T_{\rm vir}>10^{4}$K. 
The importance of HD cooling is further discussed in recent literature
in a variety of contexts, such as in relic \HII regions and in supernova remnants
of the first stars (Nagakura \& Omukai 2005; Johnson \& Bromm 2006, 2007; 
Yoshida 2006; Ripamonti 2006).
Since the formation and evolution of the first \HII regions are 
directly linked to the formation of the first stars, it is necessary
to perform a simulation starting from realistic initial conditions
under a proper cosmological set-up. The relevance, and possible importance 
of HD cooling to early star formation can be addressed only by using 
such simulations.

The overall sign of net radiative feedback effects from the first stars has 
been remaining an important, outstanding issue. 
Ricotti, Gnedin \& Shull (2002) find a net positive feedback effect by ionizing radiation
in initially over-dense regions because \HH formation is promoted by the additional 
free electrons. 
O'Shea et al (2005) also claim an overall positive feedback effect, but
its strength is still uncertain because their calculation does not include 
directly the effects of photo-evaporation.
Oh \& Haiman (2003) argue that the residual entropy 
of relic HII regions implies that gas collapses into low-mass halos with much lower 
central densities, making \HH much more vulnerable to UV photo-dissociation 
and resulting in negative feedback. In particular, they predict a relation between 
gas entropy and the strength of the LW background required to quench cooling. 
Mesinger, Bryan \& Haiman (2006) find that strong suppression of \HH production persists 
down to the lowest redshift in their simulation when a weak soft-UV radiation background 
is present. This complicated problem clearly requires self-consistent 3D 
hydrodynamics and radiative transfer. 

In the present paper, we study the formation and evolution
of early cosmological \HII regions using three-dimensional
cosmological radiation-hydrodynamics simulations. Unlike previous 
three-dimensional calculations of \HII regions that use either 
a static density field (Alvarez et al. 2006a) or do not follow 
the propagation of ionization fronts (O'Shea et al. 2005),
we couple hydrodynamics with photo-ionization heating as well as
radiative and chemical cooling self-consistently. We first locate 
a primordial star-forming cloud in a large cosmological simulation. 
We then carry out ray-tracing calculations of radiation transfer.
We show that nearby low-density gas clumps are destroyed by a 
sweeping ionization front. The ionized gas first escapes out 
of the shallow potential wells of the dark halos with a velocity 
much larger than the virial velocity, but eventually falls back
when the halos have grown large enough. Consequently, there is 
a significant time gap between the formation epoch of the first generation
star and that of the second one in the same comoving volume.
Using cosmological simulations, we follow the thermal and chemical 
evolution of the ionized gas in relic \HII regions until the gas 
recombines, cools, and recollapses at the center of the 
gravitationally growing host dark halo. We show that, at high 
redshifts ($z>10$), the minimum gas temperature of the gas cloud 
is limited by that of the cosmic microwave background
with $T_{\rm cmb} = 2.728 (1+z)$.
Because of its low temperature, the characteristic mass 
of the gas clump at first run-away collapse is found to be 
$\sim 40 M_{\odot}$. Stars formed in such gas clouds likely
have a smaller mass ($< 40 M_{\odot}$) than the
first generation of stars, and hence may be progenitors of supernovae
that enriched the gas from which recently discovered hyper metal- 
poor stars were born (Iwamoto et al. 2005; Johnson \& Bromm 2007).
Intriguingly, Tominaga et al. (2007) suggest that stars in this 
mass range trigger long-duration gamma-ray bursts with faint supernovae.

The remainder of the paper is organized as follows.
In section 2, we describe the chemistry network
employed in the cosmological simulations.
We describe our numerical techniques for radiative
transfer in section 3. 
The results from our cosmological simulations are presented 
in section 4 and section 5. The former section describes the formation
of \HII regions whereas the latter section discusses their evolution
and the formation of second-generation stars.
Finally, in section 6,
we give concluding remarks.

\section{Primordial gas chemistry}

\subsection{Hydrogen and helium chemistry}
We employ the simulation code of Yoshida et al. (2003; 2006)
that includes a non-equilibrium chemistry solver
for 14 species of hydrogen, helium and deuterium.
We use the functional fits of Hui \& Gnedin (1997)
for case B recombination rates in the present paper.
In the following, we describe in detail the relevant aspects to simulations
of (relic) \HII regions.

The rate of the charge exchange reaction
\begin{equation}
{\rm H}_2+{\rm H}^+ \rightarrow {\rm H}_2^+ +{\rm H},
\end{equation}
was recently revised by Savin et al. (2004a,b). We use this updated rate
because these processes become important when the ionization fraction of
the gas 
is large. Another important process is
collisional dissociation of \HH
by electron impacts
\begin{equation}
{\rm H}_2 + e \rightarrow 2{\rm H} + e.
\end{equation}
The commonly used values are for \HH molecules which are
initially in the $v=0$ level (e.g. Shapiro \& Kang 1987). 
Stibbe \& Tennyson (1999) show that the LTE rate, which accounts 
for dissociation from higher vibrational levels, is about two orders 
of magnitude larger. Since we are interested in cooling of gas in 
early cosmological \HII regions, where the density is typically 
less than $\sim 1 {\rm cm}^{-3}$, we assume that the higher vibrational
levels are not significantly populated and thus we use the $v=0$ rate 
in the present paper.

Next, we include charge exchange between H and He
\begin{equation}
{\rm He}^+ + {\rm H} \rightarrow {\rm He}  + {\rm H}^+, 
\end{equation}
and its reverse reaction
\begin{equation}
{\rm He} + {\rm H}^+ \rightarrow {\rm He}^+  + {\rm H}
\end{equation}
to determine the abundance of He$^{+}$ accurately.
These processes make a small but noticeable difference in He$^+$ abundance 
in a partially ionized gas at $T<5000$ K. 
%note to Peng. The difference is typically by 50\% at T<6000 K

In a low density primordial gas, the main formation path for hydrogen 
molecules is via the H$^-$ channel:
\begin{equation}
{\rm H} + {\rm e} \rightarrow {\rm H}^-  + h\nu \, ,
\end{equation}
\begin{equation}
{\rm H} + {\rm H}^- \rightarrow {\rm H}_2 + e \, ,
\end{equation}
where electrons are used as catalysts.
\HH formation via the H$_2^+$ channel
and via reactions involving HeH$^+$ ions 
are dominant only at very high-redshifts ($z>200$) where 
cosmic microwave background photons destroy H$^-$ by photo-detachment,
making the H$^-$ channel ineffective (see, e.g., Hirata \& Padmanabhan 2006).
There is considerable variation among published experimental
data and theoretical calculations for the reaction rate of
Equation (6).
Glover, Savin \& Jappsen (2006) study in detail how uncertainty in this rate 
affects the final \HH abundance, and conclude that the gas cooling
time scale is critically affected by the adopted rate. 
We use the fit by Galli \& Palla (1998) to the calculations
of Launay et al. (1991) for this rate, which is roughly intermediate
in the range of rates studied by Glover et al. (2006).

\subsection{Deuterium Chemistry}

The chemical reactions and the rate coefficients for reactions involving
deuterium are summarized in Table 1.
We note here some differences from the previous works by
Galli \& Palla (1998) and Nakamura \& Umemura (2002).
We use the updated rates of Savin (2002) for the charge transfer reaction
\begin{equation}
{\rm D}   + {\rm H}^+  \rightarrow {\rm D}^+  + {\rm H}\;\;\;\;\;({\rm reaction}\;\;{\rm D2}),
\end{equation}
and its reverse reaction 
\begin{equation}
{\rm D}^+   + {\rm H}  \rightarrow {\rm D}  + {\rm H}^+ \;\;\;\;\;({\rm reaction}\;\;{\rm D3}).
\end{equation}
For the main HD formation path
\begin{equation}
{\rm D}^+  + {\rm H}_{2}  \rightarrow {\rm H}^+  + {\rm HD}  \;\;\;\;\;({\rm reaction}\;\;{\rm D7}),
\end{equation}

\begin{inlinefigure}
\resizebox{12cm}{!}{\includegraphics{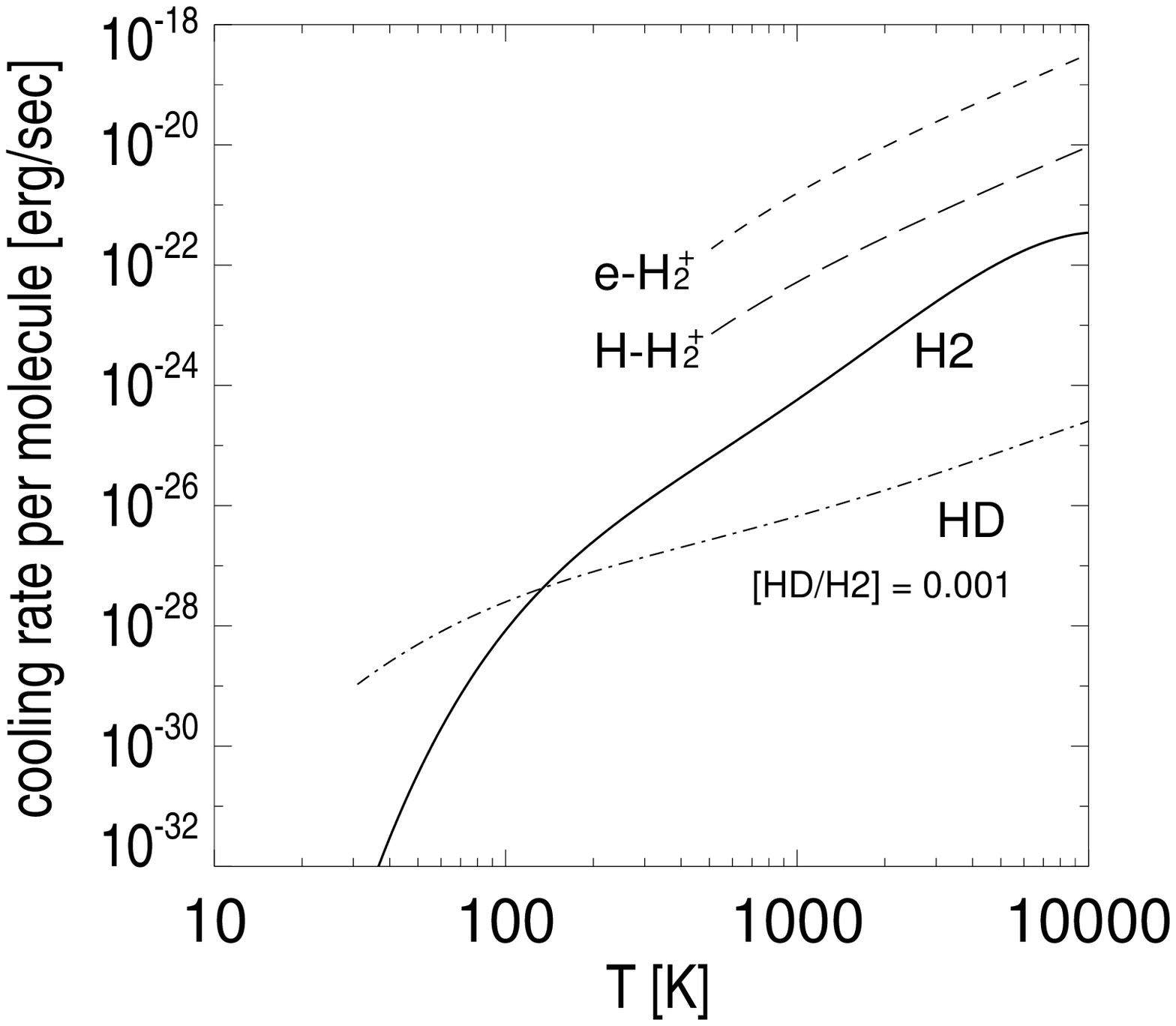}}
\caption{The molecular cooling functions for
\HH (solid), HD (dot-dashed), and \HHp (dashed).
For \HHp cooling, we show both the contributions from 
e-\HHp collisions (short dashed) and H-\HHp collisions (long dashed). 
For simplicity we assume $n_{\rm H} = n_{\rm e} = 1 {\rm cm}^{-3}$.
\label{plot1}}
\end{inlinefigure}
\noindent we adopt the rate $1.6\times 10^{-9} {\rm cm}^{3}\; {\rm s}^{-1}$ 
of Wang \& Stancil (2002)
which is slightly smaller than that in Galli \& Palla (1998).
Throughout, we set the primordial deuterium abundance 
to the standard value of $4\times 10^{-5}$. 
Reactions involving D$^-$ are unimportant 
in the regime we consider, both in a neutral
and in an ionized gas. Nevertheless we include them for completeness.

\subsection{Molecular cooling}

Radiative cooling processes owing to excitation, ionization, 
and recombination of atomic hydrogen and helium are well-determined.
We use the cooling rates of Fukugita \& Kawasaki (1994).
We also include Compton cooling because it is the dominant
cooling process in a diffuse ionized gas at high redshift.
Below, we describe the molecular cooling processes that are
important at low temperatures.

We use the cooling rate of Galli \& Palla (1998) for H$_{2}$
line cooling, and that of Flower et al. (2000) for HD line cooling
at low densities. 
Cooling by HD molecules is important only at low temperatures
($T < 200 {\rm K}$) and low densities ($n_{\rm H} < 10^8 {\rm cm}^{-3}$). 
At high temperatures, H$_2$ cooling dominates because the HD abundance 
decreases relative to \HH (see section \ref{sec:isobaric}). 
The HD cooling function of Flower et al.
does not include transitions between high vibrational levels but
the contribution to the total cooling rate at the relevant densities 
are unimportant, as shown by Lipovka et al. (2005).
HD cooling is effective at temperatures as low as $\sim 30$K.

It is well-known that radiative cooling is limited 
by the cosmic microwave background (CMB) radiation at high redshifts
because atoms and molecules act as a {\it heating} agent when $T < T_{\rm CMB}$.
The CMB temperature is given by
\begin{equation}
T_{\rm CMB}=2.73 (1+z) {\rm K},
\end{equation}
which becomes comparable to or larger than 30K at $z>10$.
Taking the CMB {\it heating} into account, we 
simply set the effective cooling rate as
\begin{equation}
\Lambda = \Lambda (T_{\rm gas}) - \Lambda (T_{\rm CMB}).
\end{equation}

Cooling by ionic molecules can be important in a gas with
a large ionization fraction, because ionic molecules are 
excited by frequent impacts with fast electrons.   
We include the cooling function of \HHp (Suchkov \& Shchekinov 1978; 
Galli \& Palla 1998). The cooling rate is given by the sum 
of two contributions (e-\HHp collisions and H-\HHp collisions), 
\begin{eqnarray}
\Lambda_{{\rm e}-{\rm H}_2^+} &=& 3.5\times 10^{-27} \exp\left(\frac{-800.0}{T}\right) T^{2}\;n_{\rm e} \\
\Lambda_{{\rm H}-{\rm H}_2^+} &=& 1.0\times 10^{-28} \exp\left(\frac{-650.0}{T}\right) T^{2}\;n_{\rm HI} \, ,
\end{eqnarray}

\begin{inlinefigure}
\resizebox{8.5cm}{8cm}{\includegraphics{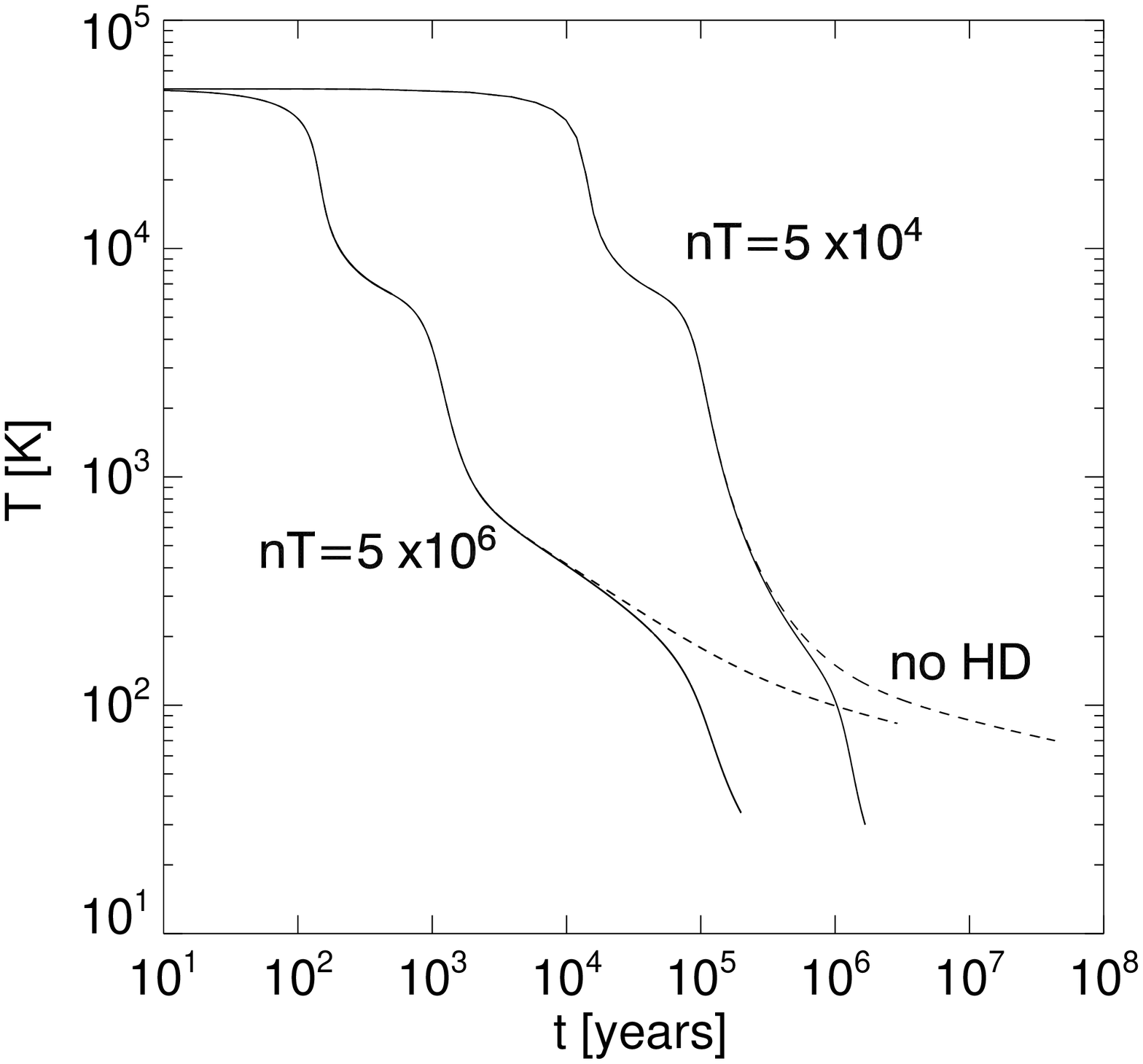}}
\resizebox{8.5cm}{8cm}{\includegraphics{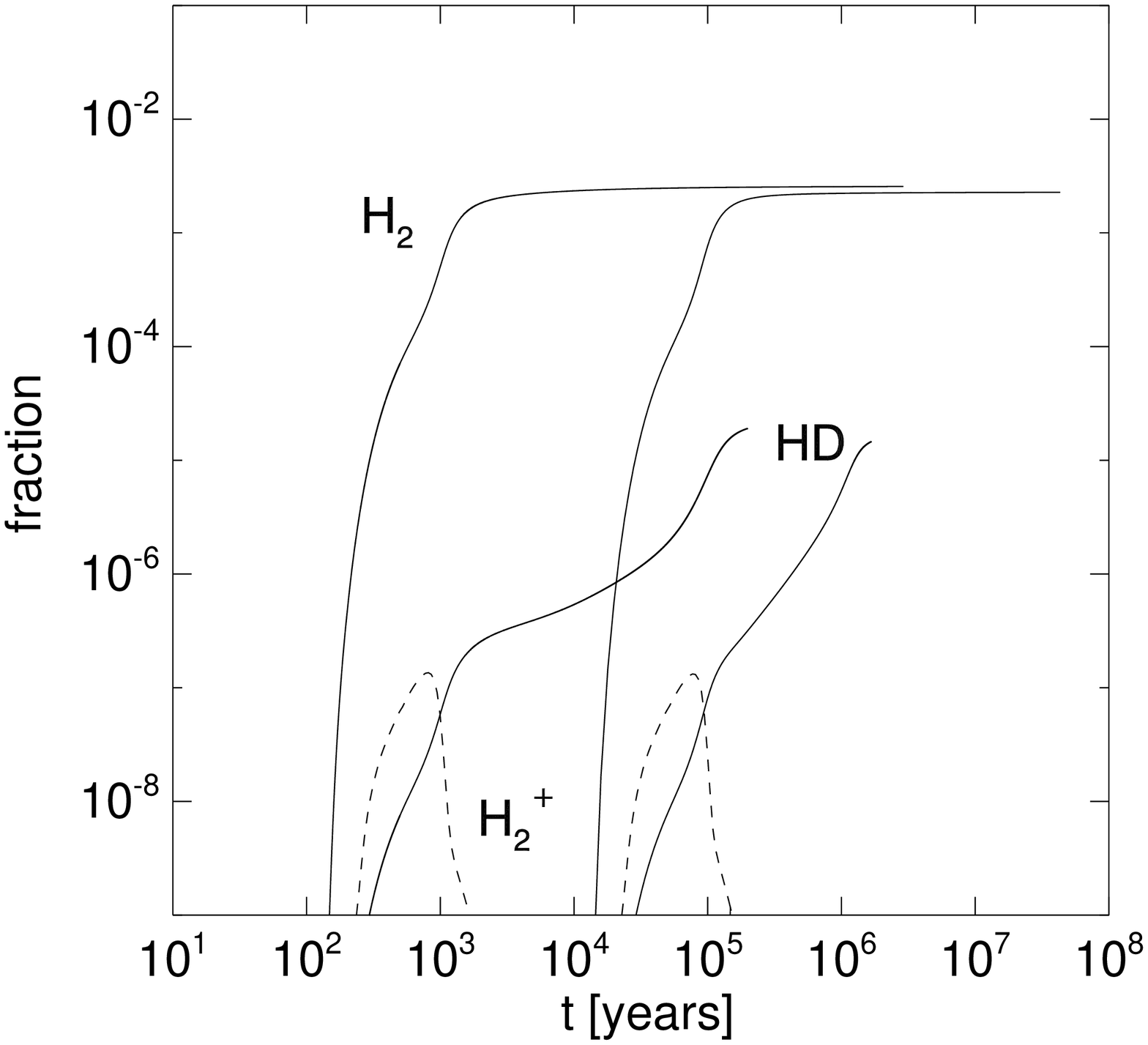}}
\caption{Evolution of the gas temperature (left) and
molecular fractions (right) for an isobarically cooling gas.
The gas is assumed to be fully ionized initially, with 
a temperature of $T=50000 K$. We run two cases with
initial densities of $n_{\rm H}=1,\;100 {\rm cm}^{-3}$.
The dashed lines in the left panel are for runs without HD cooling.
The effect of HD cooling can be seen in the temperature
and chemical evolution at $t > 10^{5}$ years. 
In the right panel, we also show the fraction of H$_2^+$ ionic
molecules.
\label{fig:isob}}
\end{inlinefigure}

\noindent where the $\Lambda$s are energy loss rates per \HHp molecule.
For the isobaric calculations we present in the next section, 
the \HHp fraction becomes as large as $\sim 10^{-7}$
when the temperature is $\sim 5000-7000$ K. Then the \HH fraction is $\sim 10^{-4}$.

We find that the above cooling processes by \HHp molecules contribute 
roughly equally to the total cooling rate as two other dominant cooling processes, 
hydrogen Lyman-$\alpha$ cooling and \HH cooling at 
temperatures $\sim 5000-7000$ K.
We also consider two other ionic molecules, H$_3^+$ and HeH$^+$.
In addition to the fiducial chemistry network, 
we include reactions for formation and destruction 
of these molecules. By calculating an isobaric test case,
we find that their fractions are always very small, and conclude
that we may ignore reactions involving H$_3^+$ and HeH$^+$.
(See Appendix for details.)

\subsection{Isobaric cooling of an initially hot, ionized gas}
\label{sec:isobaric}
Our main objective in the present paper is not only
to calculate the extent of \HII regions, but also to
follow the thermal and chemical evolution of the ionized
gas in relic \HII regions. It is intriguing that 
there is a (suggested) possibility of the formation
of low mass metal-free stars in relic \HII regions
(Nagakura \& Omukai 2005; Vasiliev \& Shchekinov 2005; 
Johnson \& Bromm 2006),
where enhanced molecular hydrogen production enables 
the gas temperature to be significantly lower than in a neutral
primordial gas cloud.

The evolution of an isobarically cooling gas serves as a 
simple, yet illustrative model of how a photo-ionized or
collisionally ionized gas evolves in collapsed halos.
Figure \ref{fig:isob} shows the temperature evolution for a parcel of gas
cooling from $T=5\times 10^{4}$ K. We assume that the gas is initially
fully ionized and then solve the full rate equations to study
the thermal and chemical evolution. 
In Figure \ref{fig:isob}, we show the evolutionary tracks for two cases,
with and without HD chemistry and cooling. 
The overall evolution except in the low temperature region appears
quite similar to that in, e.g., Oh \& Haiman (2002).
We see a clear difference in the evolution at $T<200$K, however.  
Previous calculations neglect the formation of and cooling by HD, 
and hence the gas temperature is limited to $T\sim 100$ K, at which 
point cooling by H$_{2}$ becomes
inefficient (see Fig. 1). With HD cooling, the gas further cools
down to a temperature of $\sim 30$K, where HD cooling becomes
inefficient. The right panel of Figure \ref{fig:isob} shows the evolution of the
H$_2$ and HD abundances. While the early evolutionary tracks
parallel each other, fractionation occurs at late times,
when the temperature is low, and the HD abundance is
enhanced relative to H$_2$. 
The final HD abundance is found to be about one percent of
the H$_2$ abundance, which is a factor of 200
higher than the primordial [D/H] abundance.
It is also about a factor of ten larger than
that found in standard cosmological recombination calculations
(e.g. Stancil, Lepp, \& Dalgarno 1998; Galli \& Palla 1998).

Fractionation of [HD/H$_2$] occurs for various reasons.
In the above example of an isobarically cooling gas,
one of the main formation path is the reaction
${\rm D}^+  + {\rm H}_{2}  \rightarrow {\rm H}^+  + {\rm HD}$,
for which there is no counterpart for H$_2$. 
The H$_2$ fraction in the diffuse neutral gas in the early universe is
$\sim 10^{-5}-10^{-6}$ (Stancil et al. 1998; Galli \& Palla 1998), 
whereas the universal H$_2$ fraction in the collapsing halo
is as large as a few times $10^{-3}$ (Susa et al. 1998; Oh \& Haiman 2002).
Hence, the large H$_2$ fraction promotes the formation of HD molecules at low temperatures.
In other words, HD is energetically favored because
the binding energy of HD molecules is higher than that of \HH molecules.
The relative abundance in equilibrium is
\begin{equation}
\frac{n({\rm HD})}{n({\rm H}_2)} = 2 \frac{n({\rm D})}{n({\rm H})}\;\exp\left(\frac{465 {\rm K}}{T}\right),
\label{eq:abundance_eq}
\end{equation}
and thus HD molecules are preferentially produced 
at temperatures much lower than 465 K (Solomon \& Woolf 1973). 

\begin{inlinefigure}
\vspace{2cm}
\resizebox{13cm}{!}{\includegraphics{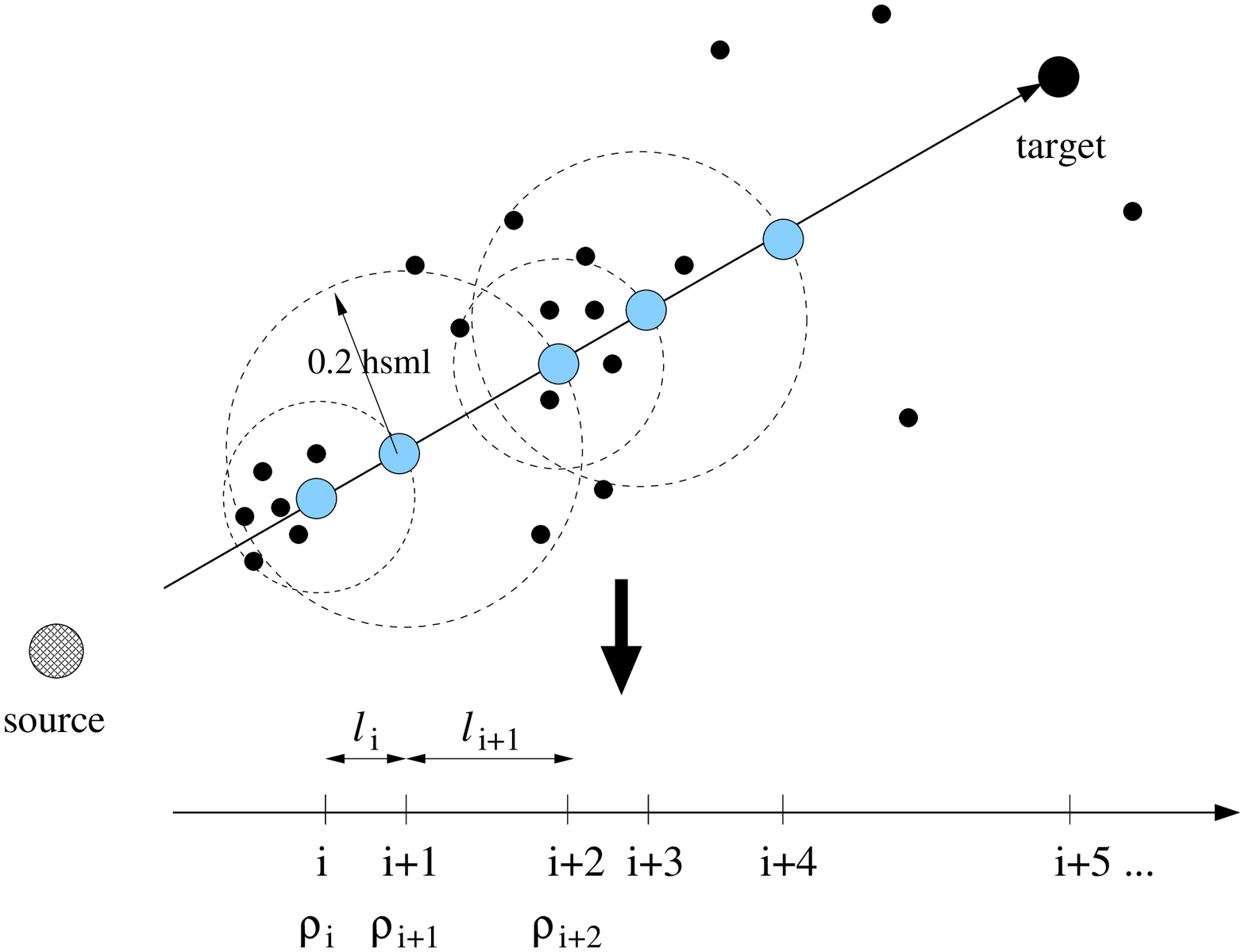}}
\caption{Schematic diagram showing how evaluation points
are defined for a given particle distribution.
The source and the target particle are connected by a straight
line. The evaluation points (large colored circles), at which the local densities
are calculated, are defined following a Smoothed Particle
Hydrodynamics procedure (i.e. the ``gather'' formalism of
Hernquist \& Katz 1989).
The spacing of two adjacent evaluation points
is given by a fraction of the local smoothing length.
For the radiative transfer calculation,
we compute the ionization-front crossing time 
using the density $\rho_i$ and the local path length $l_i$.
\label{fig:eval}}
\end{inlinefigure}

\section{Radiative transfer of ionizing photons}
In this section, we describe our numerical scheme for radiative transfer
coupled with hydrodynamics and chemistry. 
The entire procedure is fully parallelized 
and coupled with the parallel smoothed particle
hydrodynamics (SPH) code GADGET2 (Springel 2005),
so that the radiative transfer calculation, as well as the gravity and hydrodynamics,
can be performed on massively parallel architectures.
We note that the version of SPH we adopt is a fully conservative
formulation (Springel \& Hernquist 2002), where the equations
of motion properly account for evolution in the
smoothing lengths of the gas particles.
At the end of this section, we present a simple, yet very important,
test case of photo-evaporation of a small gas sphere
using our code.

\subsection{Ionization front propagation}
Our radiative transfer calculations are done in a two-step manner
as follows. For a given point radiation source, we first compute and assign photon arrival 
times for {\it all} the gas particles surrounding it. 
The photon arrival time for a gas particle is 
calculated by integrating the ionizing photon consumption 
over the path from the central source to the gas 
particle. We define a path, a photon ray, by a straight line
connecting the source and a target particle. We then compute local 
densities and smoothing lengths at many evaluation points on the path
in an SPH fashion.
Figure \ref{fig:eval} is a schematic diagram showing how 
evaluation points are defined and the local densities are calculated.
Starting from the central source, an ionization front is advanced on the 
path over the segment length $l_i$ between the $i$-th and $i+1$-th evaluation 
points. We calculate the local density $\rho_i$ and the corresponding smoothing length 
$h_i$ at the $i$-th point as in SPH. The position of the next $i+1$-th point is determined
by the condition that the length $l_i$ is smaller than the local smoothing length $h_i$;
i.e., the density variation over $l_i$ must be sufficiently 
small. We take $l_i=1/5 h_i$, after performing several tests 
to confirm that halving or doubling the numerical factor does not change 
the result.
Between two adjacent evaluation points, the I-front crossing time is computed 
by integrating the jump condition for the 
I-front position,
\begin{equation}
4\pi r_{\rm I}^{2} n \frac{{\rm d} r_{\rm I}}{{\rm d}t}=\dot{N}_{\rm ph}-4\pi \alpha_{\rm B} 
\int_{0}^{r_{\rm I}} n_{\rm e} n_{\rm p} r^{2} {\rm d}r.
\label{eq:photon_ray}
\end{equation}

\begin{inlinefigure}
\resizebox{10cm}{!}{\includegraphics{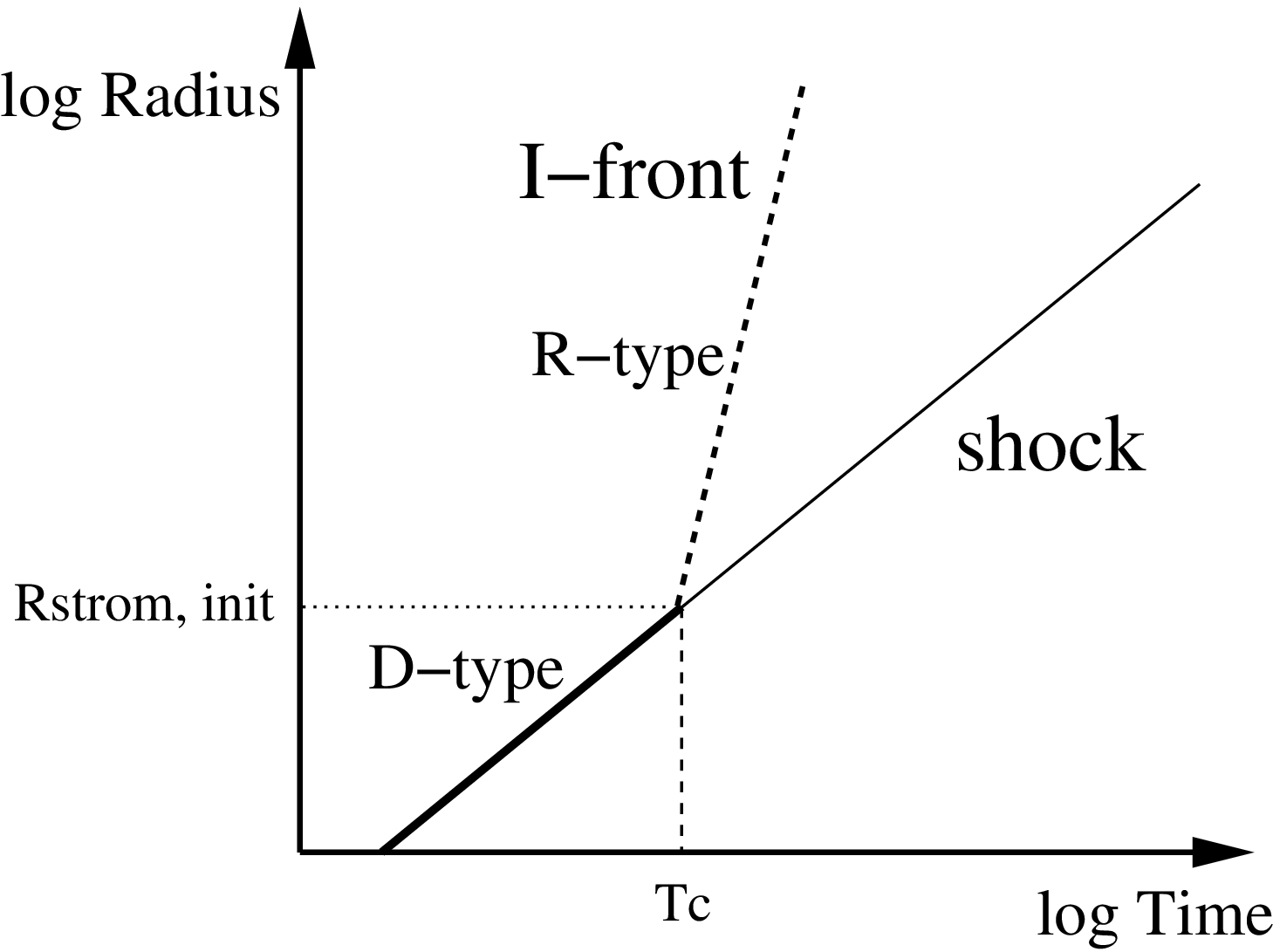}}
\caption{Characteristic time evolution
of the shock radius and the I-front radius in a gas sphere
with a steep density profile. The I-front changes from D-type to R-type
at a time $T=T_{\rm c}$. The transition point $R_{\rm strom}$ can be
evaluated from the initial Str\"omgren radius given by Equation (\ref{eq:strom}).
\label{fig:DR}}
\end{inlinefigure}

\noindent We approximate the above integral by a summation over
ray segments as
\begin{equation}
  \int_{0}^{r_{\rm I}}n_{\rm e}n_{\rm p} r^{2} dr = 
\sum_{i} n_{{\rm e}, i}n_{{\rm p}, i} r_{i}^{2} l_i \, ,
\end{equation}
assuming complete ionization for hydrogen and/or helium depending on the relative extents
of \HII and \HeIII regions
(see Section \ref{sec:helium}).
We advance the I-front by repeating this procedure until we reach the target particle,
or until the source lifetime elapses. 
In this way, every particle is assigned its own photon arrival time $t_{\rm a, i}$. 
Particles having $t_{\rm a, i} < t_{\rm life}$ 
are to be irradiated by the radiation source at $t_{\rm a, i}$.

\subsection{Treatment of D-type ionization front}
While the above method works well when I-front propagation is not 
density-bounded, i.e., of R-type, there is a situation where 
it does not provide a physically correct solution 
for $t_{\rm a, i}$ (Susa 2006; Ahn \& Shapiro 2007). 
When an I-front is of D-type, 
it can move forward only at the velocity of the foregoing
hydrodynamic shock. Namely, the I-front catches up with the
the foregoing shock as soon as the shock reduces the density
inside it (see Shu 1992 and KYSU for a detailed description of D-type
and R-type ionization fronts).
For initially steep power-law density profiles, 
$\rho \propto r^{-w}$ with $w > 1.5$, an I-front 
begins as D-type, and then becomes R-type
when the internal density is reduced by a foregoing
hydrodynamic shock that erases the initial steep density
profile. Through a series of numerical simulations, KYSU found that, 
for a given density profile,
the time when the transition of D- to R-type occurs
can be accurately estimated {\it a priori}
by a Str\"omgren analysis for the {\it initial} density profile. 
We describe the method below.

The Str\"omgren radius is obtained by equating the 
photon production rate of the central source and the
total recombination rate, 
and the result is expressed as
\begin{equation}
R_{\rm Strom} = 150 \left(\frac{\dot{N}_{\rm ion}}{10^{50} s^{-1}}\right)^{1/3}
\left(\frac{n (R_{\rm Strom})}{{\rm cm}^{-3}}\right)^{-2/3}\;\;{\rm pc},
\label{eq:strom}
\end{equation}
where $\dot{N}_{\rm ion}$ is the photon production rate of the source
and $n(R_{\rm Strom})$ is the mean density within $R_{\rm Strom}$.
The mean density $n$ within the shock decreases as it propagates and 
sweeps outward.  $R_{\rm Strom}$ increases then, and eventually the condition
$R_{\rm Strom} = R_{\rm shock}$ is met. At this time, ionization is balanced
by total recombination within $R_{\rm shock}$. This is the time when the 
I-front can {\it take off} from the shock and changes into R-type. 
Figure \ref{fig:DR} schematically shows how the shock and the I-front 
propagate. KYSU studied the time and position of the I-front transition
using one-dimensional calculations with full radiative transfer,
and showed that equation (\ref{eq:strom}) indeed provides an accurate 
estimate for the transition point.

In order to reproduce the D-to-R transition in three-dimensional
calculations, we first evaluate the initial Str\"omgren radius
using Equation (\ref{eq:strom}) for the density profile around
the source, and assume that the I-front moves at a constant shock 
velocity while it is of D-type. From the results of KYSU, we set the 
shock velocity $v_{\rm shock} = 25 {\rm km} {\rm s}^{-1}$. Although 
the exact value of $v_{\rm shock}$ is, in principle, dependent on 
the density profile and the infalling gas velocity,
it is a weak function of the density slope
over the range of gas density profiles considered in KYSU, 
and is about $25 - 35 {\rm km} {\rm s}^{-1}$ (see also Shu et al. 2002). 
Photon arrival times within $R_{\rm s}$ are simply given by the time
when the shock arrives. Once the I-front transforms to R-type, 
we can use Equation (\ref{eq:photon_ray}) to compute photon arrival 
times as described in the previous section.

\subsection{Photo-ionizationa and heating}
The second step is the calculation of photo-ionization and
photo-heating rates. Gas particles are irradiated by the central 
source after $t_{\rm a,i}$. We use an approximation that, {\it after} 
$t_{\rm a,i}$, the radiation intensity and spectrum at each particle's 
position are computed in the optically-thin limit. 
Although this assumption might seem overly simplified,
it is indeed a good approximation for the particular
problems we are interested in, where the initial gas density
profile is steep. Three-dimensional simulations 
(Abel et al. 2002; Yoshida et al. 2003, 2006) as well as  
one-dimensional calculations (Omukai \& Nishi 1998; Ripamonti 
et al. 2002) show that primordial gas clouds have
steep density profiles $\rho \propto r^{-n}$ with 
$n\sim 2-2.2$. For such profiles, only the very central part is 
dense, where ionizing photons are effectively consumed by recombinations 
initially. This dense region is, however, quickly swept by the 
hydrodynamic shock and the mean recombination rate rapidly decreases 
as well. Consequently, the opacity of the interior is progressively 
reduced and becomes less and less important as time elapses. 
Note also that stellar radiation with $13.6 {\rm eV} < h\nu < 54.4 {\rm eV}$ 
is {\it not} significantly absorbed by HI in a \HII/\HeII/\HeIII region  
for the reasons explained in section \ref{sec:helium}. 
In section \ref{sec:minihalo}, we run a test problem and show 
that the above approximation works rather well for our particular
situation.

The photo-ionization rate for hydrogen
${\rm H} + h\nu \rightarrow {\rm H}^+ + e^-$ 
is computed as
\begin{equation}
k_{\rm ion} = \int d\Omega \int^{\infty}_{\nu_{\rm L}} 
d\nu \frac{I_{\nu} \sigma_{\nu}}{h\nu},
\end{equation}
and the corresponding heating rate is given by
\begin{equation}
\Gamma = n_{\rm HI} \int d\Omega \int^{\infty}_{\nu_{\rm L}} 
d\nu \frac{I_{\nu} \sigma_{\nu}}{h\nu} (h\nu - h\nu_{\rm L}),
\end{equation}
where $h\nu_{\rm L}$ is the ionization threshold energy, 13.59 eV. 
The photo-ionization and heating rates for helium are computed 
similarly. We adopt the ionization cross sections in Osterbrock 
\& Ferland (2006). When helium ionization is included,
as described in the next section, we modify the radiation intensity
according to the ionization structure. 

\subsection{Helium ionization}
\label{sec:helium}
Massive metal-free stars emit a large number of photons with 
$h\nu > 24.5$ eV, and those with $h\nu > 54.4$ eV as well, and 
thus can ionize helium. The relative production rate of these
high energy photons (e.g., Bromm, Kudritzki \& Loeb 2001) indicates
that the structure of the ionized region around a massive 
Population III star may be similar to that of present-day planetary 
nebulae, where a small \HeIII region in the immediate vicinity 
of the star is surrounded by a larger H{\sc ii}/\HeII region. 
It is important to include helium ionization because
the temperature of the ionized gas is raised by additional 
photo-heating, and,
consequently, the pressure gradient (the driving force of 
the gas flow) is increased.
To include helium ionization, we adopt approximations
suggested by Osterbrock \& Ferland (2006):
(1) the gas is almost fully ionized within the \HeIII
region, and (2) hydrogen in the \HeIII region is kept ionized
by recombination of \HeII Lyman-$\alpha$ and Balmer 
photons as well as the continuum produced via
\HeII two-photon process.
Because the number fraction ratio of He/H is small, 
the \HII and \HeII regions have the same extent
if the source spectrum is a hot ($T\sim 10^{5}$K) thermal one
such as that for massive metal-free stars (e.g., Schaerer 2002). 
These approximations considerably simplify the remaining task.

We first compute photon-arrival times for the 
\HeIII region using Equation (\ref{eq:photon_ray}), assuming
the gas is fully (both hydrogen and helium) ionized within it.
(Note that $n_{\rm p}$ needs to be replaced by
$n_{{\rm He}^{++}}$ in equation [\ref{eq:photon_ray}]).
We then compute the extent of the \HII region surrounding
the \HeIII region following a similar procedure. 
Here, we make use of the above approximation (2),
and compute the arrival times of hydrogen ionizing photons
assuming that they are not absorbed in the \HeIII region.
In general, the boundary of an \HeIII region may not
be as sharp as we assume here, because a fraction of 
photons in the high-energy tail can penetrate further.
Nevertheless, we adopt the above simplifications to
obtain the approximate size of the \HeIII region.
Detailed structure of helium ionization will be presented
elsewhere (Kitayama et al, in preparation).
In summary, after all of these procedures, every gas
particle is assigned two photon arrival times,
$t_{{\rm a}, i, h\nu_1}$ and $t_{{\rm a}, i, h\nu_2}$
where $h\nu_1 = 13.6$ eV and $h\nu_2 = 54.4$ eV, respectively.
The particles are irradiated by photons with $h\nu > h\nu_1$ 
at $t_{{\rm a}, i, h\nu_1}$, and by those with $h\nu > h\nu_2$
at $t_{{\rm a}, i, h\nu_2}$.
The photo-heating rates are calculated consistently with this. 
Within a H{\sc ii}/\HeII region, the radiation spectrum is cut-off 
above $h\nu_2=54.4$eV, whereas within a H{\sc ii}/\HeIII region, 
we use a full thermal spectrum for the assumed hot stellar source.

\subsection{Photo-detachment and photo-dissociation}
\label{sec:shield}
We include photo-detachment of H$^-$ by photons with
$h\nu > 0.755$ eV, and photo-dissociation of H$_2$ by
Lyman-Werner (LW) photons. For photo-detachment
\begin{equation}
{\rm H}^{-} + \gamma \rightarrow {\rm H} + e,
\end{equation}
we use the cross-section of de Jong (1972)
\begin{equation}
\sigma_{{\rm H}^{-}} = 7.928\times 10^{5} (\nu-\nu_{\rm th})^{3/2} \nu^{-3}\;\;{\rm cm}^{2}
\end{equation}
for $h\nu > h\nu_{\rm th} = 0.755$eV. Since the optical depth
for these photons is generally small in the primordial intergalactic medium
(see Reed et al. 2005 for realistic values around cosmological density peaks),
we adopt the optically-thin approximation. Namely, when a radiation source is turned on, 
the radiation intensity at $0.755 {\rm eV} < h\nu < 11.18 {\rm eV}$ 
is calculated in its optically-thin limit at every point in the simulation
region. 
 
Photo-dissociation of hydrogen molecules by LW band photons
\begin{equation}
{\rm H}_{2} + \gamma \rightarrow {\rm H}_{2}^* \rightarrow {\rm H} + {\rm H},
\end{equation}
is handled in the same manner.
\footnote{Here we consider the LW photons emitted directly from the central radiation source
(a star) and not from the UV background radiation, which is considered separately
in section \ref{sec:LWbackground}.}
We use the dissociation rate
\begin{equation}
k_{\rm LW} = 1.1\times 10^{8}\;F_{\rm LW}\;\;\; {\rm s}^{-1}
\label{eq:H2diss}
\end{equation}
where $F_{\rm LW}$ is the LW photon flux 
at $h\nu = 12.87$eV in units of 
${\rm erg} {\rm s}^{-1} {\rm cm}^{-2} {\rm Hz}^{-1}$
(Abel et al. 1997). This reference photon frequency corresponds to transitions
to the $v=13$ vibrational level of the Lyman state, that lies
in the middle of the strongest transitions. 
To compute the LW radiation flux, we again adopt the optically-thin 
approximation and set it as $F_{\rm LW} \propto 1/R^{2}$ where $R$ is 
the distance from the source. Note that photo-dissociation is unimportant
within \HII regions where hydrogen molecules are destroyed mainly by 
collisional dissociation and charge transfer.
Because of the small frequency range of the \HH LW band (11.18 - 13.6 eV), 
Equation (\ref{eq:H2diss})
is a good approximation even if the radiation 
intensity varies smoothly over this range.
However, the hydrogen Lyman-series 
absorption causes a substantial intensity reduction at the line frequencies, 
some of which lie close to a few of the strongest LW lines
(Haiman, Rees \& Loeb 1997; Ricotti et al. 2002; Yoshida et al. 2003). 
We ignore the effect for computational efficiency, and hence it is expected
that photo-dissociation outside the \HII region would be
slightly less effective than reported in the present paper.

In and around dense molecular gas clouds, the effect of self-shielding 
needs to be taken into account. For a static isothermal gas, an effective 
shielding factor is given as a function of column density, by
\begin{equation}
f_{\rm shield} = {\rm min} \left[1, 
\left(\frac{N_{\rm H_2}}{10^{14} {\rm cm}^{-2}}\right)^{-3/4}\right],
\end{equation}
where $N_{\rm H_2}$ is the molecular hydrogen column density
(Drain \& Bertoldi 1996). 
We account for \HH self-shielding as in Yoshida et al. (2003)
and Glover, Savin, \& Jappsen (2006). Briefly, the radiation intensity at 
each position in the simulated region is computed by assuming that it
is attenuated through surrounding dense gas clouds.  We define a local
molecular hydrogen column density $N_{\rm H_{2}}$ in a consistent
manner employing the SPH formalism.  We line-integrate
the molecular hydrogen number density
around the $i$-th gas particle according to
\begin{equation}
N_{{\rm H_{2}}, i}=\int_{{\bf r}_{i}}^{r_{\max}} n_{\rm H_{2}} dl \, ,
\end{equation}
where ${\bf r}_{i}$ is the position of the $i$-th gas particle and
$r_{\rm max}$ is the length scale we choose in evaluating the column
density. 
In practice, we select an arbitrary line-of-sight and sum the contributions 
from neighboring gas particles within $r_{\rm max}$ by projecting an SPH spline 
kernel for all the particles whose volume intersects the sight-line. 
We repeat this procedure in six directions
along orthogonal axes centered at the position of the $i-$th particle.
We then take the minimum column density as an conservative estimate.
While the shielding effect becomes significant
for $N_{\rm H_2} \gg 10^{14} {\rm cm}^{-2}$, the gas remains nearly 
optically thin to LW photons even for column densities 
$N_{\rm H_2} \sim 10^{20-21}$cm$^{-2}$ 
if there are large velocity gradients and/or disordered motion
(Glover \& Brand 2001). 
When computing the above integral, we discard gas particles
that have a large relative velocity ($V_{\rm rel} \gg v_{\rm thermal}$)
to the $i$-th particle, in order to account for the reduction of the 
shielding effect owing to Doppler shifting, following Glover et al. (2006). 
We use a constant thermal velocity $v_{\rm thermal} = 2\;{\rm km/sec}$,
by noting that an H$_2$ rich dense gas has a low temperature of $ T \la 500$ K.
We have chosen the length scale $r_{\rm max}=100$ physical
parsec by noting that the virial radius of an early mini-halo with mass 
$10^{6} M_{\odot}$ is about 100 parsec.  There are not
significantly larger gas clumps than this scale in the simulated
region.  The local column density estimates are easily computed along
with other SPH variables with a small number of additional operations.

 We treat photo-dissociation of HD molecules similarly. 
Since some of the HD Lyman-bands lie close to those of \HH,
the HD lines can be shielded by \HH. Barsuhn (1977) argues
that the HD dissociation rate can be decreased by as much as a factor of
three. Since HD molecules are formed in H$_2$-rich regions,
we assume that the shielding by H$_2$ is maximally effective,
and adopt this reduction factor for HD photo-dissociation.

\begin{inlinefigure}
\resizebox{16cm}{!}{\includegraphics{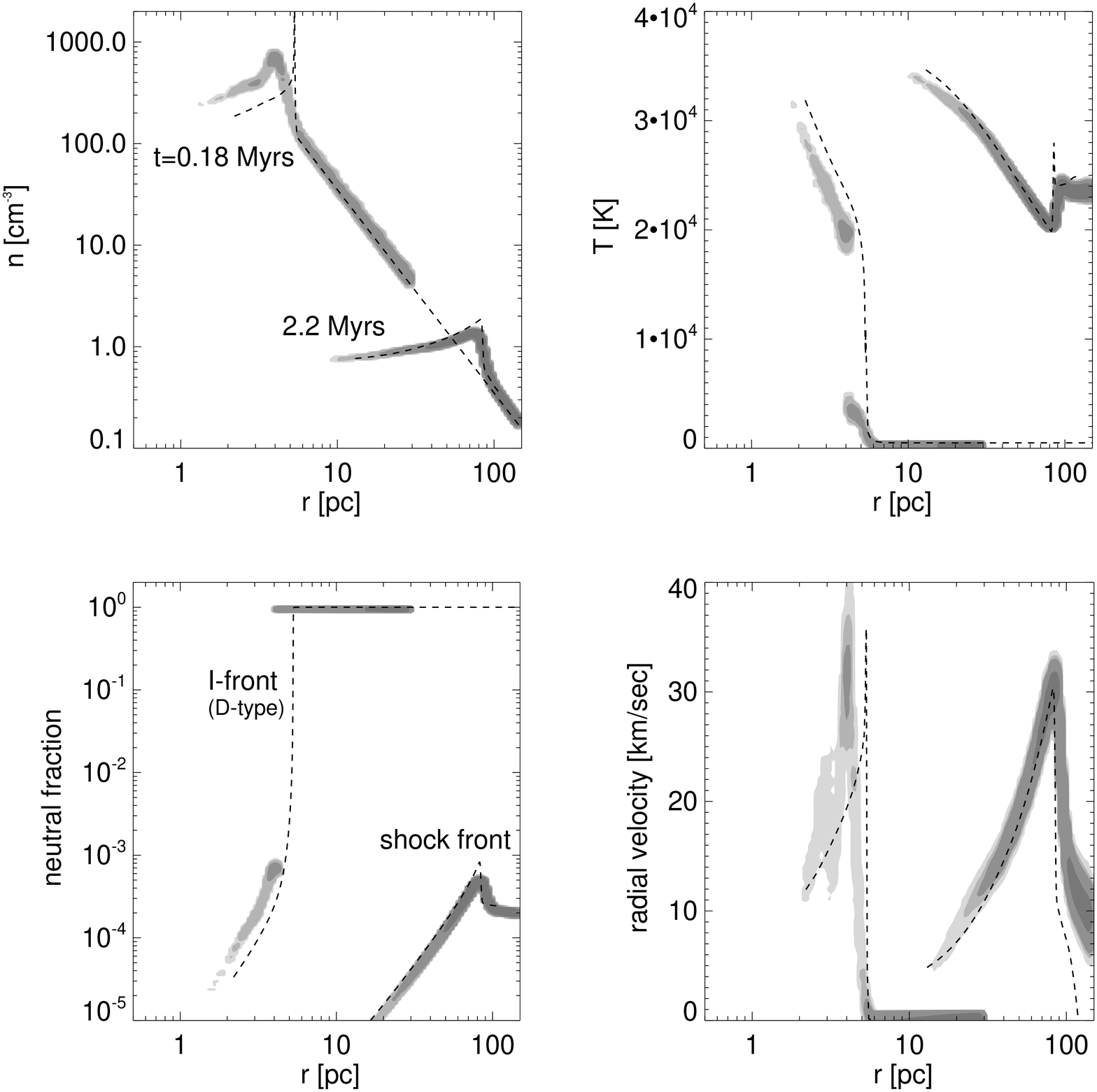}}
\caption{Minihalo evaporation test problem.
We plot the radial profiles of density (top-left), 
temperature (top-right),
neutral hydrogen fraction (bottom-left), 
and radial velocity (bottom-right)
at two output times, t=$1.8\times 10^{5}$ yrs 
and $t=2.2\times 10^{6}$ yrs.
We compare the results of our three-dimensional calculation 
(shown as grey contours) with those of the full-radiative 
calculation of KYSU (dashed lines).
\label{fig:HII}}
\end{inlinefigure}

\subsection{A test problem: Mini-halo evaporation}
\label{sec:minihalo}
We test the overall accuracy of our radiation hydrodynamics calculation
using a spherical cloud problem.
Expansion of an I-front in a spherical gas cloud
offers a simple, yet illustrative case for this purpose.
We adopt a spherical gas cloud problem studied by KYSU:
a massive Population III star with 
$M_*=200 M_{\odot}$ is embedded at the center of a $10^{6} M_{\odot}$
dark halo that is collapsing at $z=20$. We set the initial gas density profile 
to follow a power law $n = 3000 (r/{\rm pc})^{-2} {\rm cm}^{-3}$.
We distribute 0.2 million particles in the sphere according to this
density profile. The source is chosen to have a thermal spectrum
with effective temperature $T_{\rm eff}=10^{5}$ K. For this
test problem we include only hydrogen. 
(See KYSU for further details.)

Figure \ref{fig:HII} shows the evolution of the \HII region
around the central star.
We compare the radial profiles of density (top-left),
temperature (top-right), neutral hydrogen fraction (bottom-left),
and gas velocity (bottom-right) with those of KYSU. 
While there is a slight off-set in the position
of the D-type I-front at $t=0.18$ Myrs (note that the difference is 
{\it only} $\sim$1 pc), the agreement 
between the profiles is very good 
both in amplitude and shape. The off-set likely owes to a slight 
difference between the assumed velocity of the D-type front and the
actual shock velocity. Although we can easily fix the velocity
of the D-type I-front so that the position of the D-type front is perfectly matched
to the result of the 1D calculation at a given epoch, we do not
attempt to do so here because the difference is only transient,
and the initial behavior of the I-front does not much affect the final 
results for the long-term evolution of \HII regions. 
In Figure \ref{fig:HII}, the sharp transition at $r \sim 4$pc ($t=0.18$ Myrs) 
is a D-type front, which changes to R-type at $t\sim 0.2$ Myrs.
The profiles at $t=2.2$ Myrs show a clear bump at $r\sim 100$ pc 
that is a shock front.
The R-type ionization front is located about 1 kpc
away from the center, so the entire region is ionized
by this time, as can be seen in the neutral fraction profile.
At $t=2.2$ Myrs, the radial velocity peaks at $\sim$30 km/sec,
which is also well reproduced.

From Figure \ref{fig:HII}, we conclude that the accuracy of our code is 
quite satisfactory. The transition from D-type to R-type 
ionization front as well as hydrodynamic variables such as 
density and temperature are accurately reproduced in the 3-D 
calculation. Furthermore, even the neutral fraction profile
is reproduced rather well. We emphasize that it is generally easy 
to obtain the correct
ionized fraction within \HII regions because it is close to
unity, whereas it is hard to produce a correct neutral fraction
because it usually has a small value and is sensitive
to the local density, temperature, and their time evolution.
Hence, the level of agreement with the accurate solutions shown 
is reassuring. Our radiation transfer scheme is similar to 
that of Susa (2006) and the accuracy of our code to the particular 
problem of minihalo evaporation appears as good as Susa's result.
We mention that Susa's code can be applied to more general problems, 
because it updates gas opacities more frequently as density structures 
evolve. The method described in this section is suitable for 
radiative transfer problems where there are physically separated point
sources which have short lifetimes, such as those for early \HII regions. 
Further numerical implementations
will be necessary for more general problems.

\section{Cosmological Simulation}
We can discuss the long-term evolution of early \HII/\HeIII regions 
only if they are studied in a fully cosmological context.
To this end, we run a cosmological simulation adopting the concordance 
$\Lambda$CDM cosmology with matter density $\Omega_{m}=0.26$, 
baryon density $\Omega_{\rm b}=0.04$,
cosmological constant $\Omega_{\Lambda}=0.7$ and expansion rate at the
present time $H_{0}=70$ km s$^{-1}$Mpc$^{-1}$. The density fluctuation amplitude
is normalized by setting $\sigma_8 = 0.9$. While some of these cosmological
parameters are different from those determined most recently from 
the third-year WMAP data (Spergel et al. 2006), all of our results presented 
below are robust to slight changes in the values. Nevertheless, it is probably
important to note how adopting a smaller value of $\sigma_8$ 
would affect
the epoch of the events such as first-star formation.
Generally, the formation epoch is shifted to lower redshifts
for lower values of $\sigma_8$. For the value from 
the third-year data,
$\sigma_8 = 0.74$, everything occurs later, about 40\% in 
cosmological expansion parameter (e.g., Alvarez et al. 2006b).
Hence, if the first stars are formed at $z \sim 20$ in the
old $\Lambda$CDM model with $\sigma_8 = 0.9$, then it would be
at $z\sim 12-13$ in models with $\sigma_8 = 0.74$.
This may be important when we consider the evolution 
of gas in relic \HII regions. The temperature
of the cosmic microwave background radiation, which is a function of
redshift, will affect the evolution of the prestellar gas cloud
collapsing in the relic \HII region. We discuss this issue in section 
\ref{sec:second}. We also note that the fact that structure formation occurs late 
in low $\sigma_8$ models may actually provide better prospects for 
direct observations of the first stars and early \HII regions.

We start from a low-resolution simulation of boxsize $1.4$Mpc 
on a side. We identify one of the most massive halos in the simulation
volume at a redshift of 20, and then make zoomed initial conditions
around this object with a much
higher mass resolution (see Fig. \ref{fig:panel}).
In the highest resolution region,
the gas particle mass is $1.82 M_{\odot}$ and the dark matter
particle mass is $11.8 M_{\odot}$.
We use the parallel Tree $N$-body/SPH solver GADGET2 (Springel 2005)
to evolve the system from $z=100$ to the epoch 
when the first collapsed gas cloud is identified.
The code includes all the relevant atomic and molecular processes
as described in Yoshida et al. (2003, 2006).
Figure \ref{fig:panel} shows the simulated high-resolution 
region and a boundary low-resolution region.
 We locate a cold ($T< 500$ K) and dense ($n > 10^3\;{\rm cm}^{-3}$)
cloud whose mass exceeds $300 M_{\odot}$ as a site for star-formation.
The critical mass is obtained from simulations of Abel et al. (2002) 
and Yoshida et al. (2006) for gas clouds at the first run-away collapse.
The dark matter halo which hosts the star-forming cloud has
a mass of $5\times 10^{5}M_{\odot}$.
We define the halo virial radius such that the mean
mass density within the radius is 200 times the critical density
and the halo mass is the mass within the virial radius.
We embed a 100$M_{\odot}$ PopIII star 
at the densest part of the gas cloud and switch on radiation. 
The stellar mass is derived from the recent work of Yoshida et al. (2006)
that follows the proto-stellar evolution in detail.
For the source spectrum,
we use a blackbody with effective temperature $T_{\rm eff}$
which is determined from the stellar mass (Schaerer 2002).

We do not trigger a supernova explosion in the simulations
in the present paper for several reasons.
First, the fate of Population III stars
with mass $\sim 20-100 M_{\odot}$ is rather uncertain in terms
of their mechanical feedback efficiency (Umeda, Nomoto \& Nakamura 2000; 
Heger \& Woosley 2002), while very massive stars
with $\sim 140-260 M_{\odot}$ likely cause complete
destruction of the surrounding structure (Bromm, Yoshida \& Hernquist 2003).
Kitayama \& Yoshida (2005) show that the halo destruction efficiency 
of supernova explosions is sensitively dependent on the explosion energy,
initial density of the explosion site and even on reshift, because
of efficient Compton cooling in the early Universe.
Therefore, it is necessary to use {\it outputs} of simulations
of \HII regions, such as those presented in the present paper,
for the study of supernova feedback effect.
We will defer a consideration of these effects to future work.

\subsection{Formation of the First HII Region}
\label{sec:formation}
%this section largely edited by naoki
Figure \ref{fig:panel} shows the extent of the \HII region
around the first star at $t=$1.1, 3 Myrs. 
The \HII region has a few kilo-parsec diameter at its maximum extent,
and the ionized gas has a temperature of
$\sim 1-3 \times 10^{4}$ K (see Fig. \ref{fig:profiles}).
The diameter is similar to what has been found in previous works
on \HII regions around massive primordial stars (KYSU;
Alvarez et al. 2006a; Abel, Wise \& Bryan 2006).
The \HII region has a characteristic ``butterfly'' shape
which occurs because fast recombination in dense regions such as
filaments prevents the I-front from propagating. 
Although it is hard to see in the figure, a close look
reveals that I-front propagation is also delayed by forming gas clumps
near the main halo. However, since the maximum density of such
small gas clumps is still $\sim 1\;{\rm cm}^{-3}$, they are
quickly photo-evaporated within the central star's lifetime
(Shapiro, Iliev, \& Raga 2004; Ahn \& Shapiro 2007; Susa \& Umemura 2006).

\begin{inlinefigure}
\resizebox{15cm}{!}{\includegraphics{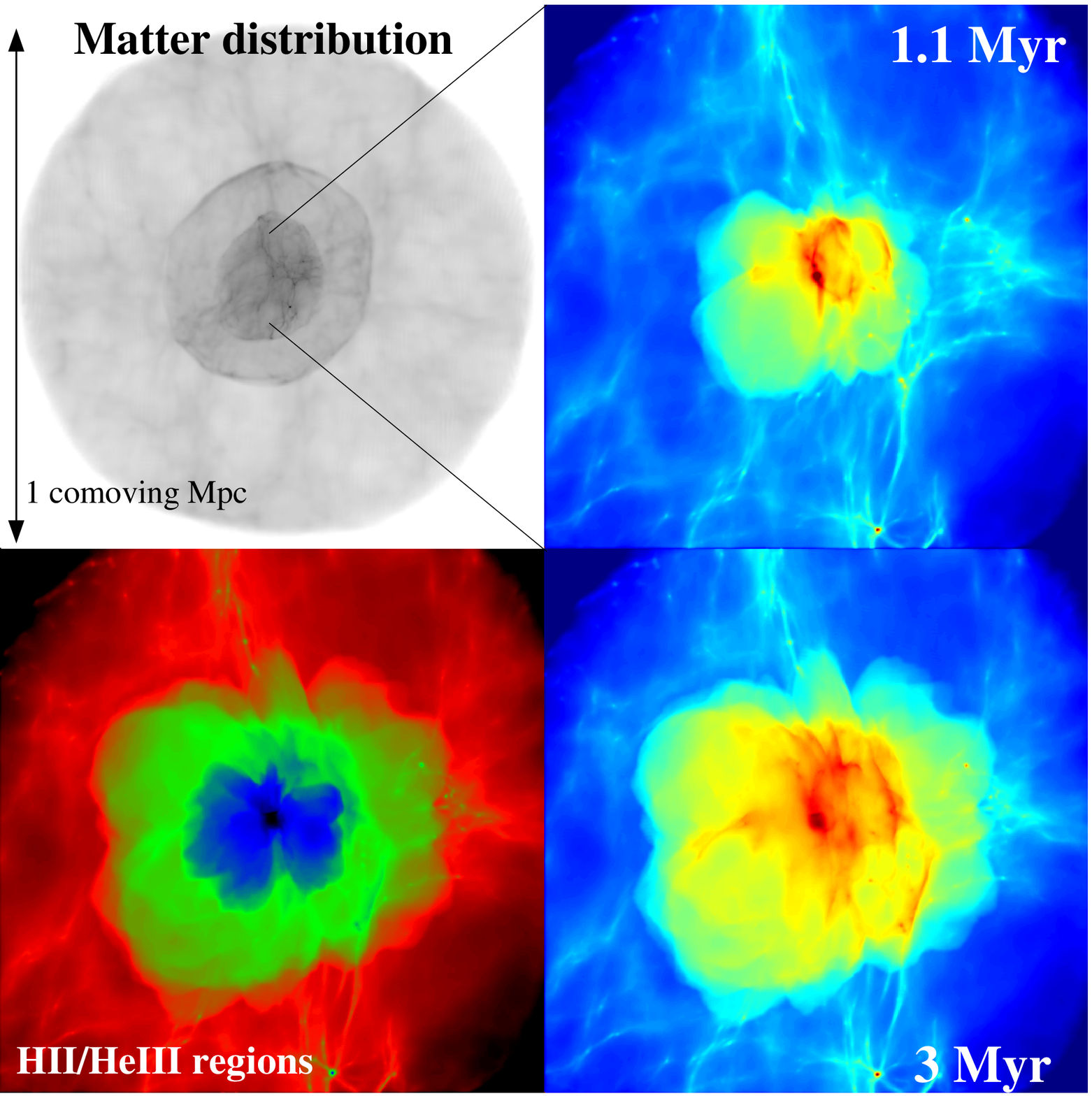}}
\caption{Expansion of the HII region around the first star.
The top-left panel shows the large-scale density distribution 
around the first object at $z=26$. A three-level zoom-in technique 
was used to make the initial conditions for this simulation. 
Panels on the right show the projected gas density weighted by ionization 
fraction. Each panel has a side-length of about 7 proper kpc.
We show the extension of the \HII region at t=1.1, 3 Myrs after the
central star turns on, as indicated in the figure. 
The bottom-left panel shows the extent of the \HII region (green)
and that of the \HeIII region (blue) at $t=3$ Myrs. 
We use a different weighting and color scale for this panel.}
\label{fig:panel}
\end{inlinefigure}

\noindent This is because there is only one halo in which the gas has cooled
by \HH cooling at $z=26$. Other surrounding halos have masses smaller than 
$\sim 10^{5} M_{\odot}$, which is much smaller than the critical collapse
mass of $\sim 5\times 10^{5} M_{\odot}$.

Since the gas in the main halo had initially a steep density gradient,
(almost) prompt photo-ionization by the R-type front 
results in a significant pressure gradient across the \HII
region, and the thermal pressure drives a strong gas wind outwards.
The darkest round part at the center of each panel on the right in 
Figure \ref{fig:panel}
is the region where the outgoing shock is evacuating the gas out of the host halo.
We show the detailed structure of the ionized region in the bottom-left
panel. We see a smaller \HeIII region, of diameter $\sim 1$ kpc (colored in blue), 
surrounded by a larger \HII (and \HeII) region. Both hydrogen and helium
in the main halo gas are nearly fully ionized, as can be seen from the extent
of the \HeIII region. Within the \HeIII region, the gas has a temperature of
$\sim 3 \times 10^{4}$ K.

Figure \ref{fig:profiles} shows the radial profiles
for density, velocity, and temperature at $t=0$ and $t=3$ Myrs. 
Thermal pressure quickly suppresses the steep 
density profile and turns it to a nearly flat one within the stellar lifetime.
The ionized region has a high temperature of $T\sim 30000$K
in the case with helium. For reference, we also show the temperature
profile from a run in which we include only hydrogen.
Additional heating by helium ionization is clearly seen.
Throughout the evolution, the peak gas velocity is about 30 km/sec, 
which is nearly 10 times larger than the escape velocity of the
host halo. Hence it is expected that almost all the gas
will be evacuated within a few tens of million years
(for a halo with virial radius $\sim 100$ pc).

\begin{inlinefigure}
\resizebox{9cm}{!}{\includegraphics{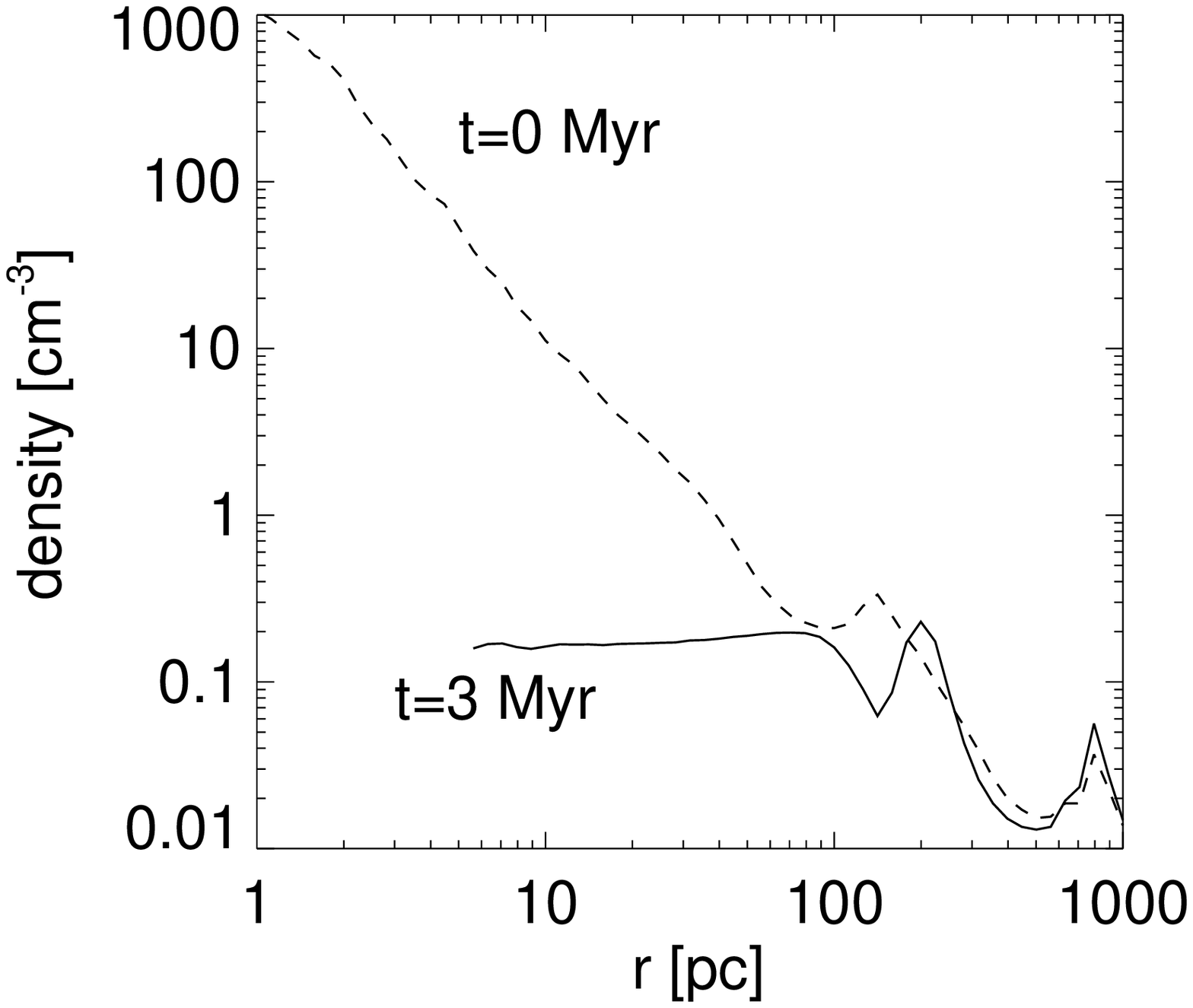}}\\
\resizebox{9cm}{!}{\includegraphics{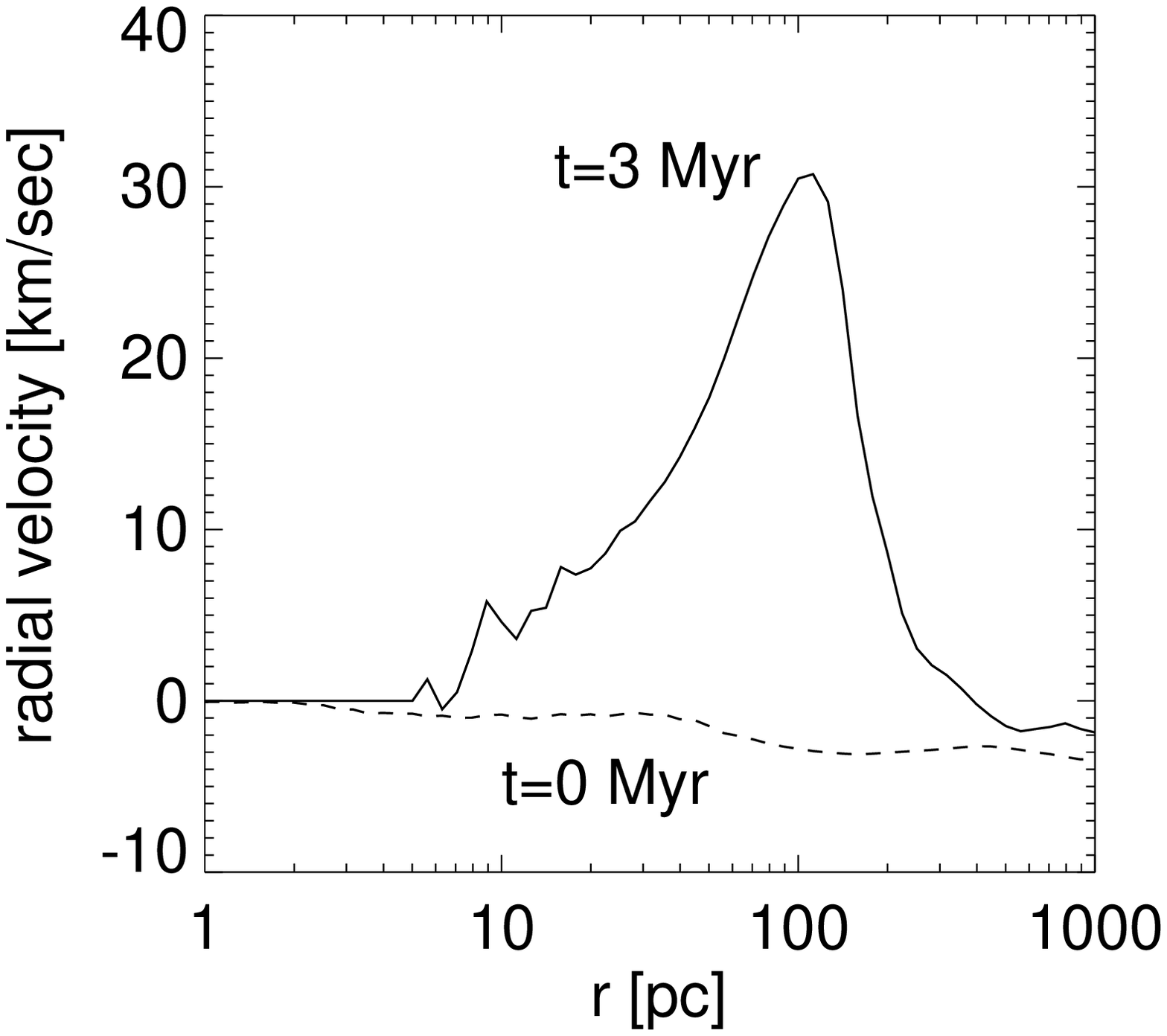}}\\
\resizebox{9cm}{!}{\includegraphics{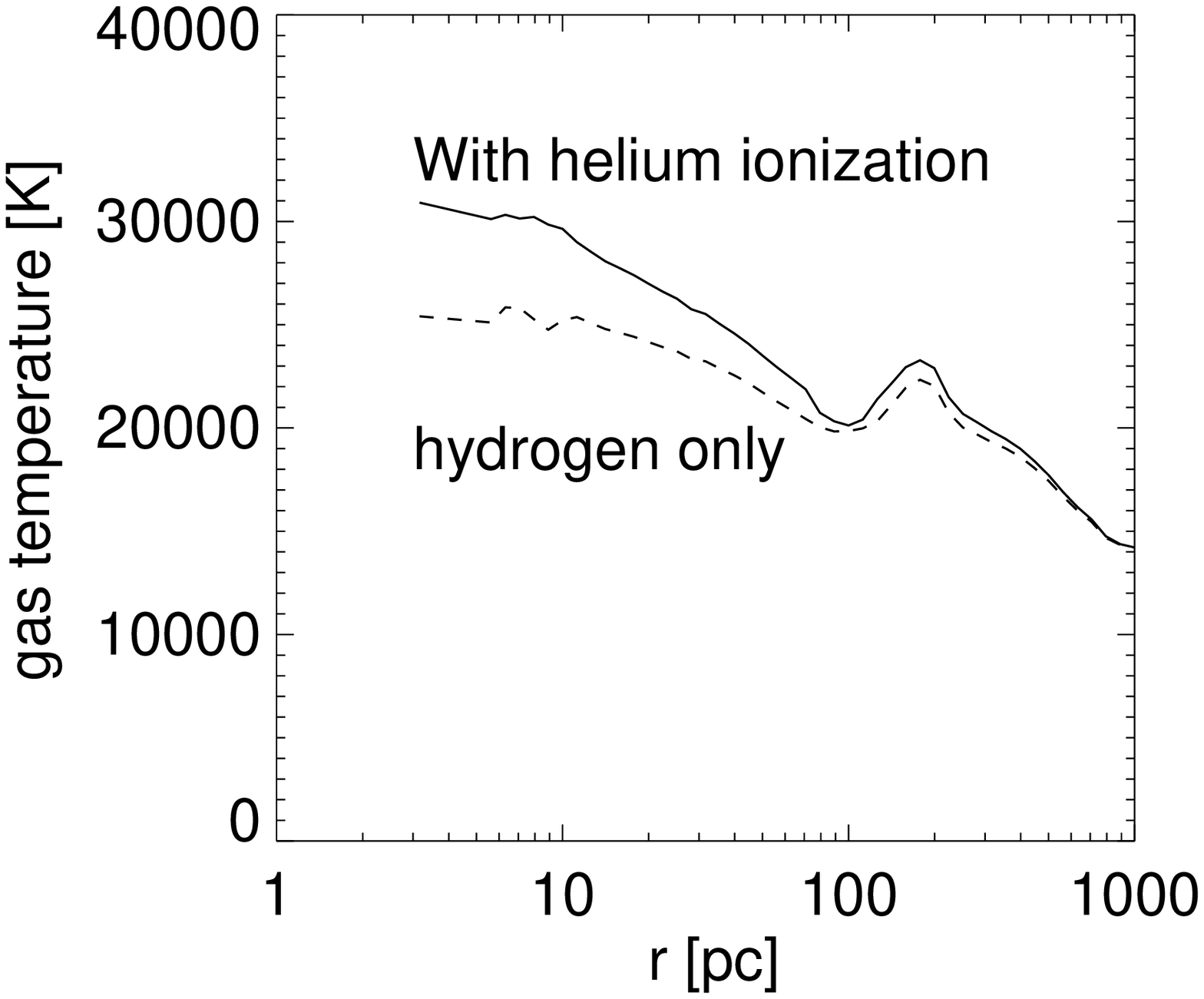}}
\caption{Radial profiles of density (top), velocity (middle),
and temperature (bottom) at $t=0$ and $t=3$ Myrs. For the temperature 
profile, we show the results of two cases: one with helium ionization
and the other with hydrogen only. (See text in section \ref{sec:formation}.)}
\label{fig:profiles}
\end{inlinefigure}

\begin{inlinefigure}
\resizebox{12cm}{!}{\includegraphics{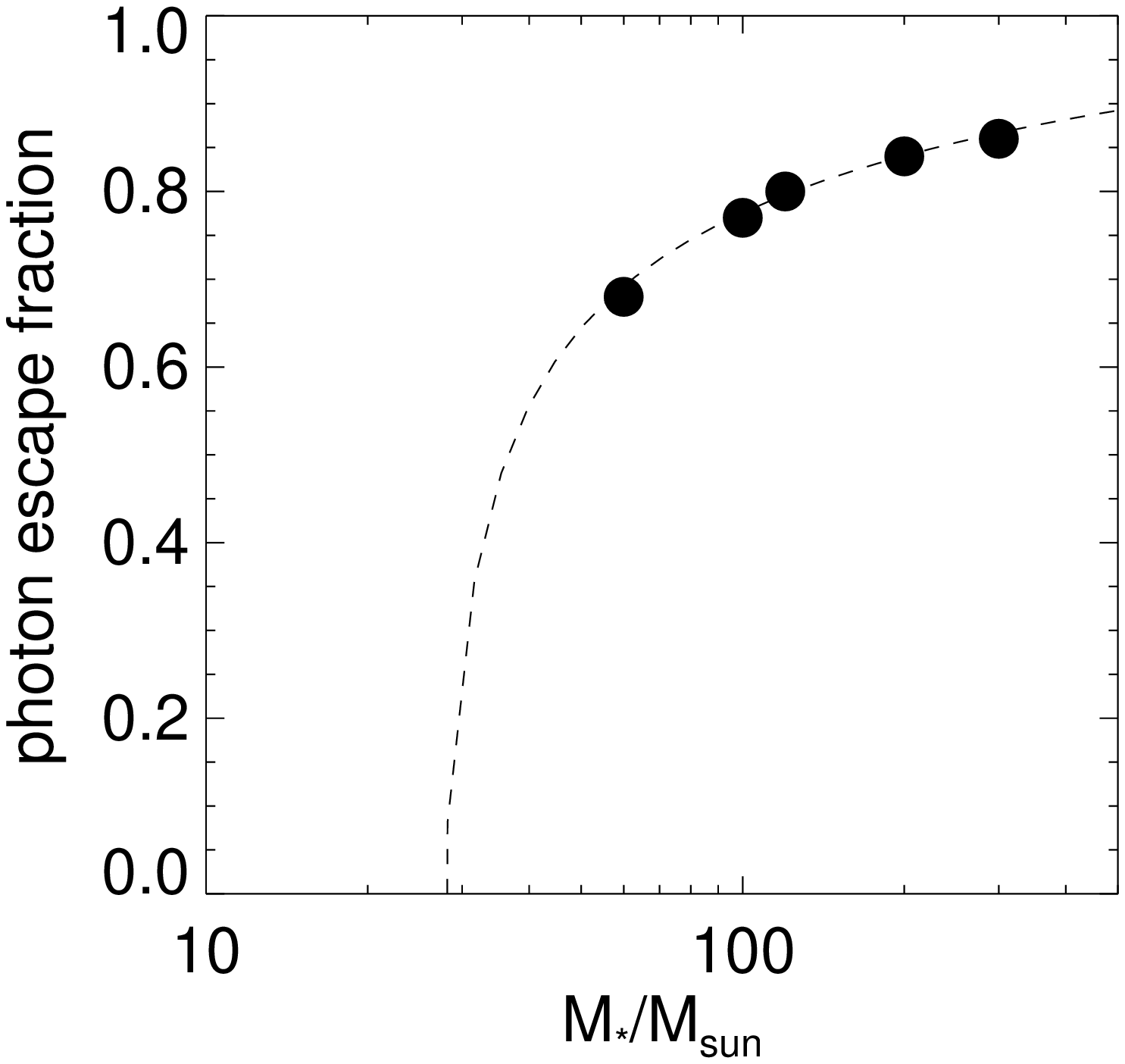}}
\caption{Time-averaged ionizing photon escape fraction
as a function of stellar mass. 
\label{fig:fesc}}
\end{inlinefigure}
\noindent Because of this efficient evacuation by radiation, further star-formation
in the same halo is quenched for a long time until
the gas can cool, be re-incorporated, {\it and} condense again. 
We study the evolution of the gas in the relic \HII region
further in the next section.

 An important quantity often used in simulations and theoretical models
of reionization is the ionizing photon escape fraction.
Our three-dimensional radiation hydrodynamics calculation 
allows to directly measure this quantity.
Since the mass resolution of our simulations is
sufficient to resolve the smallest gas clumps
within early mini-halos, the effect of gas clumping
is self-consistently included in our calculations. 
Therefore, {\it we do not need
to assume the so-called photon escape fraction nor gas clumping
factor}. The photon escape fraction is indeed obtained as a {\it result} 
of our simulations.
We define the escape fraction as 
the number fraction of photons that escape
out of the host halo's virial radius
into the intergalactic medium.
Following Alvarez et al. (2006a),
we evaluate the escape fraction along a photon ray as
\begin{equation}
f_{\rm esc}(t) = 1 - \frac{4\pi \alpha_{\rm B}}{\dot{N}_{\rm ph}}\int_{0}^{r_{\rm vir}} 
n^{2} (r,t)\;r^{2}\; {\rm d}r \, ,
\label{eq:esc}
\end{equation}
where $r_{\rm vir}$ is the virial radius of the
host halo. The halo has $r_{\rm vir}=80$ pc at $z=26$.
We compute the escape fraction averaged over the lifetime of the star.
The value obtained is $\bar{f}_{\rm esc} = 0.77$ for our fiducial model with $M_{*}=100 M_{\odot}$.
Figure \ref{fig:fesc} show the ionizing photon
escape fraction as a function of the central 
stellar mass. The result
is consistent with the results of previous 
one-dimensional calculations assuming spherical symmetry (KYSU). 
We calculate a few other cases by varying the mass of the star. 
Figure \ref{fig:fesc} shows $\bar{f}_{\rm esc}$ as a function of the stellar mass.
We find that the escape fraction as a function of the stellar mass
can be well-fitted by
\begin{equation}
\bar{f}_{\rm esc} = 2.0 - \exp\left(\frac{1.0}{(M_* - 25.0)^{0.37}}\right),
\end{equation}
for the particular host halo we simulate, i.e, a halo with mass $5\times 10^{5} M_{\odot}$
at $z=26$. 
The obtained escape fraction is in excellent agreement with that of 
Alvarez et al. (2006a);
we have confirmed the result with a $\sim$70 times better mass resolution.
The dependence of the escape fraction on halo mass and collapse
epoch, gas density profile etc. can be easily inferred following the
description of D-to-R type transition of the I-front as discussed by
KYSU.

\begin{inlinefigure}
\vspace{1cm}
\resizebox{16cm}{!}{\includegraphics{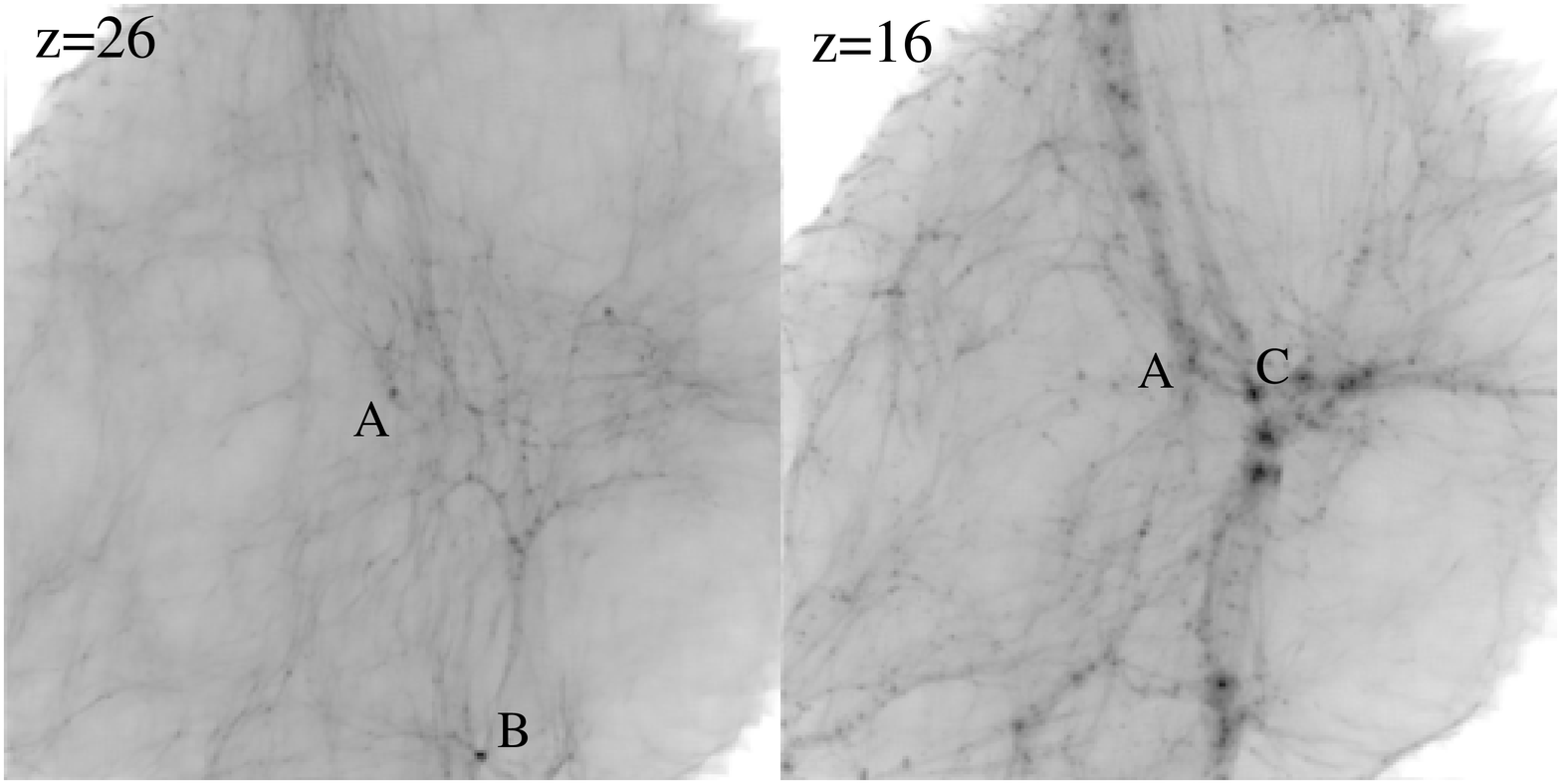}}
\caption{Underlying dark matter density field at $z=26$ 
when the first star is formed (left), and at $z=16$ when the second-generation
star(s) are formed (right), showing a significant evolution in $\sim 100$ Myrs.
The main halo is marked by A. Halo B is a nearby massive
halo, whose evolution is studied in section \ref{sec:nearby}.
The descendant of Halo A is indicated in the right panel,
where there is another halo (C) close to it. Halo C hosts
the second generation star-forming gas cloud which 
was formed from an ionized gas.
\label{fig:dm_evo}}
\end{inlinefigure}

\section{Evolution of relic \HII regions and their surroundings}
In this section, we study the thermal evolution and re-collapse of 
the gas in relic \HII regions. We first discuss possible star-formation
in regions outside the \HII region during and after the central star's lifetime, 
and then show the results for the chemo-thermal evolution of the ionized gas.
We clarify how the gas cloud evolution
differs from that for primordial gas clouds 
(e.g., Yoshida et al. 2003, 2006).

\subsection{Evolution of Nearby Halos}
\label{sec:nearby}
There are many low-mass halos in the regions surrounding the first star, 
where ionizing radiation does not arrive within the lifetime of the 
central star. 
Although the photo-dissociation region (PDR) extends much farther 
than the \HII region, the central massive star is short-lived
and thus the effect of photo-dissociation is only temporal.
We first study the impact of radiation from the first star only,
ignoring the effect of a global cosmic UV background radiation field.
After the first star dies, some of the nearby halos are able to 
re-form enough hydrogen molecules to meet the baryonic cooling criteria
(Tegmark et al. 1997; Yoshida et al. 2003; Oh \& Haiman 2003). 
Throughout the evolution, photo-dissociation of \HH, 
photo-detachment of H$^{-}$ and \HH molecule formation act 
in a complex manner. The Lyman-Werner photons continuously destroy
hydrogen molecules within a large region, and
photons with energy greater than 0.755eV effectively destroy \Hminus
(see section \ref{sec:shield}).
However, the effect is weak in regions sufficiently distant
from the radiation source, where collisional detachment processes
still regulate the abundance of \Hminus\ just as in the case without
radiation.
During the lifetime of the first star, the outer PDR remains \HH poor but 
the \Hminus\ abundance is {\it not} greatly reduced, owing to the difference
in their respective cross-sections.
The gas in distant mini-halos has a high temperature
($\sim T_{\rm vir}$) because of the absence of an effective coolant.
Hence, as soon as photo-dissociating radiation
is switched-off when the central star dies, \HH molecules are quickly 
reformed. It is interesting that the reformation takes place
in a synchronized way, determined by the death of the first star.
This \HH regeneration mechanism differs from 
the often claimed positive feedback, in which
\HH molecules are quickly formed in a (partially) ionized region
behind the I-front (Ricotti, Gnedin \& Shull 2002)
or behind a M-type shock (Ahn \& Shapiro 2007; Susa \& Umemura 2006)

In our simulation, there is a `second' halo at $\sim 2$ kpc away,
in which the gas cools and condenses after the first star dies
(see Halo B in Fig. \ref{fig:dm_evo}). Interestingly, Halo B has
a slightly larger mass than the main halo (Halo A) at $z=26$,
but the growth of Halo B has been so rapid that significant dynamical
heating associated with progenitor mergers prevented the gas
in Halo B from cooling, as often found in simulations of early structure
formation (Yoshida et al. 2003; Reed et al. 2005; O'Shea \& Norman 2007). 
Hence the first star is formed earlier in Halo A.
At the position of Halo B, the dissociation (owing to the stellar 
radiation from Halo A) timescale is
just about 1 Myrs (from equation [\ref{eq:H2diss}]). 
The \HH fraction first slightly decreases, but rapid re-formation of \HH
leads to efficient cooling and condensation within a few million years
after the first star dies.
Thus star-formation will take place in Halo B on a similar timescale, 
just as for Halo A, and the formed star, if massive, will produce a 
large \HII region. We have carried out ray-tracing calculations assuming 
that a 100 $M_{\odot}$
star is formed in Halo B. We find a similarly large \HII region,
merging (percolating) with the first \HII region.

\begin{inlinefigure}
\resizebox{12cm}{!}{\includegraphics{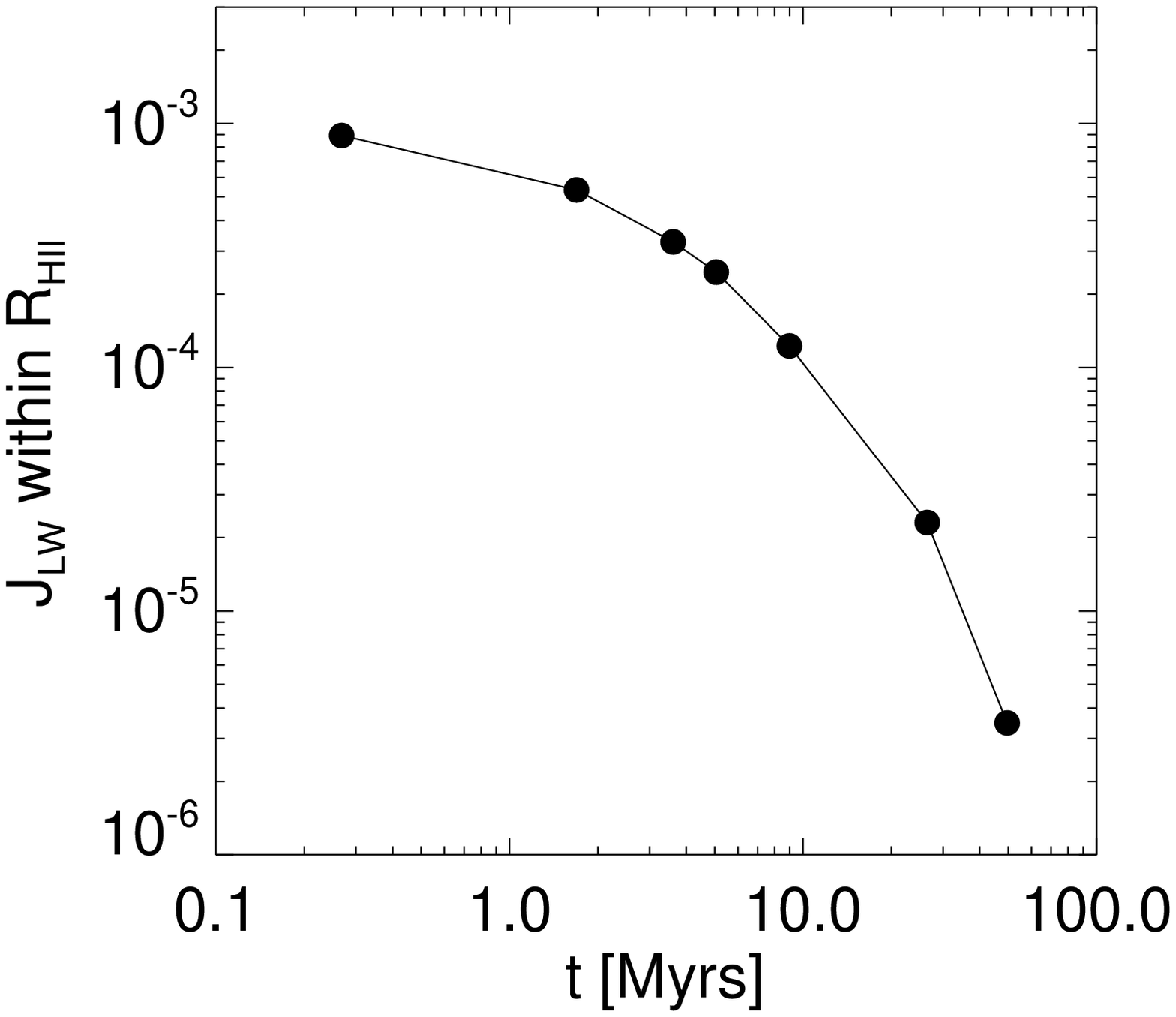}}
\caption{Evolution of the radiation intensity in 
the Lyman-Werner band, $J_{\rm LW}$, in units of
$10^{-21} {\rm erg}\;{\rm s}^{-1} {\rm cm}^{-2} 
{\rm str}^{-1} {\rm Hz}^{-1}$. The corresponding
photo-dissociation time is approximately 
$\sim 100\;(J_{\rm LW}/10^{-4})^{-1} \;\;{\rm Myrs}$.
We consider recombination photons from \HeI two-photon
and \HeII two-photon decay.}
\label{fig:helium_rec}
\end{inlinefigure}

\noindent It would be very interesting to follow the large-scale
gas dynamics in this \HII complex, but the limited simulation
volume does not allow us to study the effect of other
minihalos that possibly reside further out.
  
We note that the timing of the star-formation in nearby halos, 
and whether or not it can actually happen, is very sensitive
to details of the configuration. In the particular simulated region, 
there happened to be another halo (Halo B) just collapsing when the first 
star is being turned on, at a `favorable' location which is just outside the 
first \HII region. The survival of \HH molecules in such halos and the 
halo gas itself under the influence of radiation from the first star depends 
on the luminosity of the first star, distance from it, and the
gas density and molecular fraction profile in the halo at an exactly
corresponding time. Therefore, it is not straightforward to determine 
the fate of nearby halos in general.
Recently, Susa \& Umemura (2006) and Ahn \& Shapiro (2007) addressed
this issue using a number of simulations that explore a large 
parameter space. It appears that the process is indeed very complicated
and is simply case-dependent. A `typical' case may be found
by running a large set of cosmological simulations such as those
presented in the present paper for many halos in various environments.

\subsection{Formation of second-generation objects in relic \HII regions}
\label{sec:second}
In this section, we study the thermal and chemical evolution of
a gas cloud which is re-collapsing gravitationally from a once-ionized
gas in the relic \HII region.
To do so, we deliberately restrict ourselves to the vicinity of the first star;
i.e., we ignore (possible) gas cooling and star-formation occurring in 
places remote from the \HII region.
We also run a simulation without \HII region calculation, 
in order to make a direct comparison with runs with various
radiative feedback effects described in the following section.

\subsubsection{Photo-dissociation by recombination photons}
We first consider photo-dissociation of hydrogen molecules
by recombination photons from within the \HII region itself.
The gas with ionized hydrogen and helium emits photons in a 
broad energy range by recombination processes.
The dominant contribution to the Lyman-Werner band (from 11.2-13.6 eV) 
involves two-photon processes from helium atoms and ions. 
Since \HeII is a hydrogenic ion, we can use parametric fits 
for the two photon profile (Nussbaumer \& Schmutz 1984); 
we find that each \HeII $2s \rightarrow 1s$ decay produces 
0.07 LW photons. We now need to find the fraction of recombinations 
that result in two-photon decay from the 2s level, 
$f=\alpha_{\rm eff(2s)}/\alpha_{\rm total}$.
We can again use the fact that \HeII is a hydrogen-like ion, 
so that the effective recombination coefficients obey 
$\alpha_{\rm eff}(Z,T)=Z \alpha_{\rm eff}(1,T/Z^{2})$. Using the 
coefficients $\alpha_{\rm eff}$ for HI tabulated in
Osterbrock \& Ferland (2006), we find that at $T=20,000$ K,
$f=0.2$, and thus that $\sim 0.014$ LW photons are produced
per \HeII recombination. The corresponding contribution from 
the $2s \rightarrow 1s$ transition in \HeI is more difficult 
to calculate since \HeI is not hydrogen-like, but it is roughly 
comparable.

\begin{inlinefigure}
\resizebox{12cm}{!}{\includegraphics{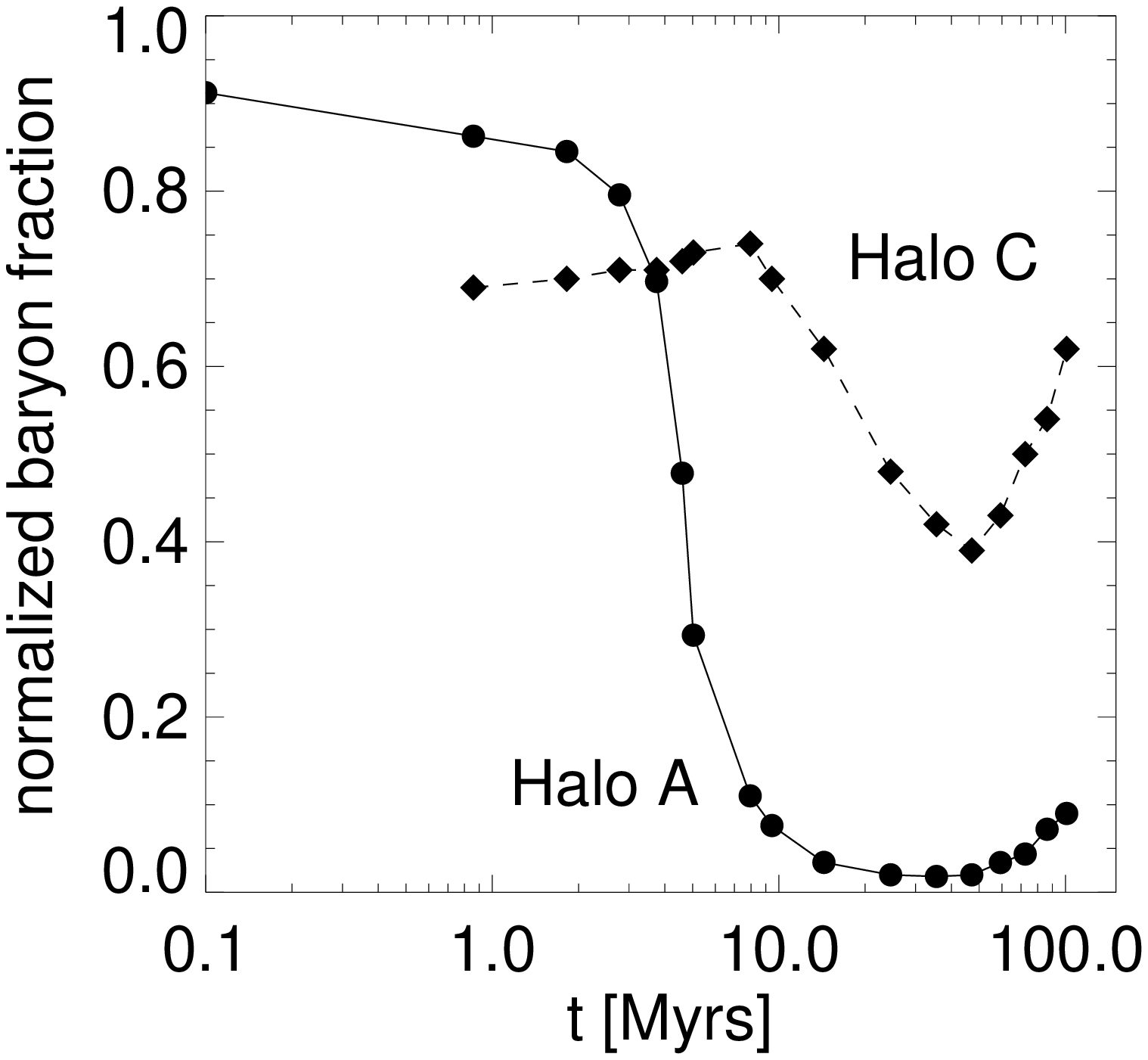}}
\caption{Evolution of the baryon fraction
of the main halo (Halo A) and a nearby `second-generation' halo
(Halo C) from $t=0$ to $t=100$ Myrs. 
We define the baryon fraction as 
$f_{\rm b}= M_{\rm gas}/M_{\rm tot}$ where
$M_{\rm tot}$ increases as the halo grows.}
\label{fig:baryonfrac}
\end{inlinefigure}

We use directly the simulation outputs to calculate the production rate of LW photons
in the relic \HII region as a function of time.
Ideally, one would like to obtain the intensity and the spectral shape
of the diffuse recombination radiation in the LW band at every point
within the relic \HII region. This would involve a very costly radiative 
transfer calculation, whereas it is sufficient for our purposes
to make a rough estimate of the importance of the recombination radiation in terms
of molecule destruction, as is done in, e.g., Johnson \& Bromm (2007).
To this end, we make the following assumptions.
We assume that the emitted LW photons are instantaneously re-distributed within 
a spherical region of radius of 1 kpc (approximately the size of the \HII
region at its maximum extent). From the production rate
of LW photons, $\dot{N}_{\rm LW}$, we approximate the intensity as
\begin{equation}
J_{\rm LW} \sim 2.2\; \left(\frac{\dot{N}_{\rm LW}}{10^{50} {\rm s}^{-1}}\right)
\end{equation} 
in units of $10^{-21} {\rm erg}\;{\rm s}^{-1} {\rm cm}^{-2} 
{\rm str}^{-1} {\rm Hz}^{-1}$. Note this estimate scales with the
size of the assumed spherical region as $J_{\rm LW} \propto 1/(4\pi R^{2})$.

Figure \ref{fig:helium_rec} shows the evolution of $J_{\rm LW}$
as a function of time elapsed since the central star died.
The maximum (at the beginning) is just about $J_{\rm LW} \sim 10^{-3}$,
and decreases slowly over several tens of million years.
The large relic \HII region continues to expand, both by thermal pressure
and by cosmological expansion,
so the density in the outer part decreases, and
helium recombination occurs progressively more slowly.

It is clear that photodissociation by LW photons with this low flux
is negligible (see equation [\ref{eq:H2diss}] and also Machacek et al. 2001;
Yoshida et al. 2003). 
As is shown in the next section, a part of the ionized gas is re-incorporated 
by a large mass halo in about 50 million years, when $J_{\rm LW} < 10^{-5}$,
and thus we can safely ignore the effect of recombination radiation
on the formation of second-generation objects in the present context.

\subsubsection{Gas evacuation and re-incorporation}
\label{sec:secondstar}
The fate of the out-flowing gas is of considerable interest,
because it determines whether or not star-formation can
be triggered promptly after the epoch of first star-formation.
In other words, the gas cooling efficiency (and hence star-formation rate)
in higher mass systems that are formed later on is critically
dependent on how efficiently the halo gas is re-incorporated.
As we show below, recollapse
of the ionized gas occurs in a very complex manner.
First, hierarchical assembly of dark matter halos is itself a complex process,
and it is often found that the descendant of an early massive halo
at one epoch is not the most massive one at a later epoch
(e.g. Yoshida et al. 2003; Gao et al. 2005).
We find exactly such a case in our simulation, where the main halo
at $z=26$, which is the host of the first star-forming cloud,
grows rather slowly, and a few other smaller halos (at $z=26$) grow 
more quickly.
After the halo gas is evacuated by the first star, 
cooling and condensation of the gas takes place at the center
of {\it another} halo that is located $\sim 0.5$ kpc away from 
the main halo at $z=16$ (see the right panel of Fig. \ref{fig:dm_evo}).

In order to see clearly when this re-incorporation happens, 
we measure the evolution of the baryon fraction of
the halo. The result is shown in Figure \ref{fig:baryonfrac},
where we normalize the baryon fraction
by the global mean $\Omega_{\rm b}/\Omega_{\rm m}$.
We show the evolution of baryon fraction
for two halos: one is the main halo and its descendant,
and the other is a different halo (Halo C in Fig. \ref{fig:dm_evo})
in which a second-generation 
object is formed later.
Initially, the gas in the main halo is evacuated by radiation, 
with the baryon fraction
rapidly decreasing from $t\sim 2$ Myrs when the hydrodynamic shock reaches
the halo virial radius.
The baryon fraction decreases down to a few percent(!) at $\sim 40$ Myrs,
and starts increasing rather slowly. 
By then the halo has already grown more than twice in mass. 
As the gas cools, thermal pressure is lost and the gas eventually falls
back, but full reincorporation is not achieved even after 
100 Myrs. We thus conclude that the first stars, if they are massive, impact
their surroundings by decoupling the gas distribution 
from that of dark matter for as long as a Hubble time at
$z\sim 20$. The descendant of the main halo (Halo A) still has a very low
baryon fraction of $\sim$ 10\% at that time.

These results have interesting implications for the growth
of putative Population III remnant black holes. Because of the rapid evacuation
of the gas around a massive star, it is likely that
the remnant black hole, even if it is formed by the death
of the star, will not be able to accrete the surrounding
gas efficiently. We defer a detailed study of this process
to future work (Li et al., in preparation), but note briefly
the physical conditions around the first star remnant.
We mark a dark matter particle (a ``black hole'') that is 
located closest to the position of the central star. 
We then keep track of the position of the ``black hole''
and calculate the mean density and temperature
with a sphere of 10 physical parsec radius around it.
We find that the local density is $\sim 0.2\;{\rm cm}^{-3}$ in the
beginning (see Fig. \ref{fig:profiles}), and
remains at $\sim 0.02-0.2\;{\rm cm}^{-3}$ 
for over 50 million years. With this low density,
the rate of gas accretion is expected to be very small.

The strength of radiative feedback is further appreciated
by the following experiment.
We ran an additional simulation in which we include no radiative feedback
from the first star.
Namely, we continued the original simulation with \HH cooling
to see if star-formation can proceed if there is no feedback.
\footnote{We terminated the run at $z=18$ because extremely small time steps are
required for gas particles in cold dense gas blobs, which effectively stops
the run. However, outputs by $z=18$ are enough to make necessary comparisons.}
The normalized baryon fraction for Halo A in this run remains roughly constant,
at $f_{\rm b, norm} = 0.9$,  
and the amout of cold $(T < 500{\rm K})$, dense 
($n_{\rm H} > 10^{3} {\rm cm}^{-3}$) gas
increases from 300 $M_{\odot}$ at $z=26$ to 
more than $10,=000 M_{\odot}$ by $z=18$.
Clearly, a large amount of cold gas should be available for further 
star-formation if there is no radiative feedback from the first star.
We thus conclude that the net feedback effect from the first
star is negative in terms of the efficiency of star-formation in the same location.

We now return to the issue of the formation of the so-called second-generation 
stars. The host of the second-generation object, Halo C, 
grows quickly and so it re-captures the evaporated gas faster than Halo A does.
Halo C has many progenitors at $z=26$, which are seen as small dark matter
clumps along a few filaments in the lower-right region of Halo A in 
Figure \ref{fig:dm_evo}.
Merging of these numerous clumps builds up Halo C at $z=16$ as is seen
in the right panel of \ref{fig:dm_evo}. 
The initial ($t<10$ Myrs) behavior of the baryon fraction for Halo C 
shown in Figure \ref{fig:baryonfrac} may be misleading because
there are many progenitors of Halo C at this epoch.  
Simply, the main progenitor is not well-defined then, and thus the 
baryon fraction for Halo C at $t<10$ Myrs may be regarded for a halo 
that is one of the main progenitors of Halo C.
 It is also worth mentioning that many of the progenitors of Halo
C are not within the \HII region; i.e., the gas within them is not
ionized. Neutral and ionized gases are mixed when Halo C is assembled,
although the central part of Halo C consists mostly of ionized gas.

The re-captured gas starts cooling and condensing towards the center of 
Halo C, forming the so-called second-generation object.
We followed the evolution of the second-generation object
using the technique of Yoshida et al. (2006) until 
the central density reaches $\sim 10^{15} {\rm cm}^{-3}$.
We emphasize that the maximum density achieved here is many orders of 
magnitude greater than previous works which study the evolution
of cooling gas in \HII regions. We are, for the first time, able to 
study details of the structure and evolution of the `cosmological' 
second-generation star-forming gas cloud.
Figure \ref{fig:HDpanel} shows the radial profiles around this object
for density, temperature, molecular fraction (\HH and HD), 
and the ratio of enclosed gas mass to the locally estimated Bonnor-Ebert mass.

The overall features appear similar to primordial neutral gas clouds 
(see, e.g., Fig. 3 of Yoshida et al. 2006). An important difference is the minimum
temperature of $\sim 50$ K which is limited by $T_{\rm CMB} = 46$ K at $z=16$
in the present calculation. The thermal evolution of the recollapsing gas
is notably different from the primordial case
in that \HH cooling is now more effective,
owing to the enhanced molecular fraction. 
This brings the gas temperature below $\sim 150$ K,
where HD cooling becomes important (see section \ref{sec:isobaric}).
We compare the temperature profile with that of the primordial case
of Yoshida et al. (2006). The difference in the minimum temperature
and the corresponding mass scale is clearly seen. 
Efficient HD cooling causes the temperature to bottom out 
at $\sim 50$ K ($\sim T_{\rm CMB}$) 
at a mass scale of 100 $M_{\odot}$. There, the HD fraction 
increases to $\sim 2\times 10^{-5}$, as is seen in the left-lower
panel in Figure \ref{fig:HDpanel}. 
In the low temperature regime, 
the relative \HH and HD abundances approach their equilibrium values
(see equation[\ref{eq:abundance_eq}]).
The temperature increases toward the center
within a mass scale of $\sim 100 M_{\odot}$, 
and the HD fraction temporarily decreases following approximately 
the equilibrium abundance (note the exponential factor  
$\exp(465 {\rm K}/T)$ in equation [\ref{eq:abundance_eq}].)
The density profile is close to a power law $n \propto r^{-\alpha}$
with $\alpha \sim 2.4$ depending on the distance range.
The profile is slightly steeper than in the primordial gas clouds
simulated by Abel et al. (2002), Yoshida et al. (2006)
and Gao et al. (2006).
Because of the lower minimum temperature owing to HD cooling,  
the temperature increase toward the center is large, by a factor of 50,
and the effective equation of state in the collapsing gas 
$M_{\rm enclosed} \la 100 M_{\odot}, n_{\rm H} > 10^{4} {\rm cm}^{-3}$,
is harder than the neutral primordial case, $P = K \rho^{\gamma}$ with
$\gamma \sim 1.2-1.3$ if expressed polytropically. 
From the Larson-Penston similarity solution for a polytropic
gas, the density gradient may be estimated as (Larson 1969)
\begin{equation}
\frac{\partial \ln \rho}{\partial \ln r} = \frac{-2}{2-\gamma},
\end{equation}
which is $\sim 2.5$ for $\gamma \sim 1.2$, in good agreement
with the actual density gradient.
The temperature increases toward the center
within a mass scale of $\sim 10 M_{\odot}$, 
and the HD fraction temporarily decreases following approximately 
the equilibrium abundance. 
However, {\it both} \HH and HD fractions increase again when three-body reactions 
convert almost all the hydrogen atoms into molecules,
making the cloud core of $\sim 1 M_{\odot}$ fully molecular. 
The behavior of the abundances of \HH and HD almost parallel
each other in this regime.

To see the onset of run-away collapse, we compare 
the enclosed gas mass with the locally estimated 
Bonnor-Ebert mass (Bonnor 1956; Ebert 1955):
\begin{eqnarray}
M_{\rm BE} &=& \frac{m_{1} c_{\rm s}^{4}}{G^{3/2} P_o^{1/2}},\nonumber \\
&\approx &20 M_{\odot} T^{3/2} n^{-1/2} \mu^{-2} \gamma^{2} \, ,
\label{eq:BEmass}
\end{eqnarray}
where $m_1$ is the first maximum mass of the solution for the isothermal 
Lane-Emden equation (see, e.g. Stahler \& Palla 2004), and $\mu$
and $\gamma$
denote the mean molecular weight and adiabatic index, respectively.
We approximate the external pressure by its local value
taken from the radial density and temperature profiles.
(Note that $M_{\rm BE}$ evaluated in this manner is essentially
the same as the local Jeans mass.)

We find that the enclosed gas mass exceeds the
Bonnor-Ebert mass at a mass scale of $M \sim 40 M_{\odot}$. 
This is the characteristic mass of the collapsing gas cloud.
The so-called second-generation object formed in this process
may be a formation site of low-mass primordial stars 
(Mackey, Bromm \& Hernquist 2003; Johnson \& Bromm 2006).
The mass of such stars and the possible difference 
from the first stars are of particular importance.
Until the time of the last output, when the central density
is $\sim 10^{15} {\rm cm}^{-3}$, we observe no sign
of fragmentation of the pre-stellar cloud, and thus
it is likely that a single proto-stellar seed is
formed in this object.
Detailed calculations of proto-stellar evolution 
for this object will be presented elsewhere
(Yoshida, Omukai \& Hernquist, in prepartion).

\begin{inlinefigure}
\resizebox{17cm}{!}{\includegraphics{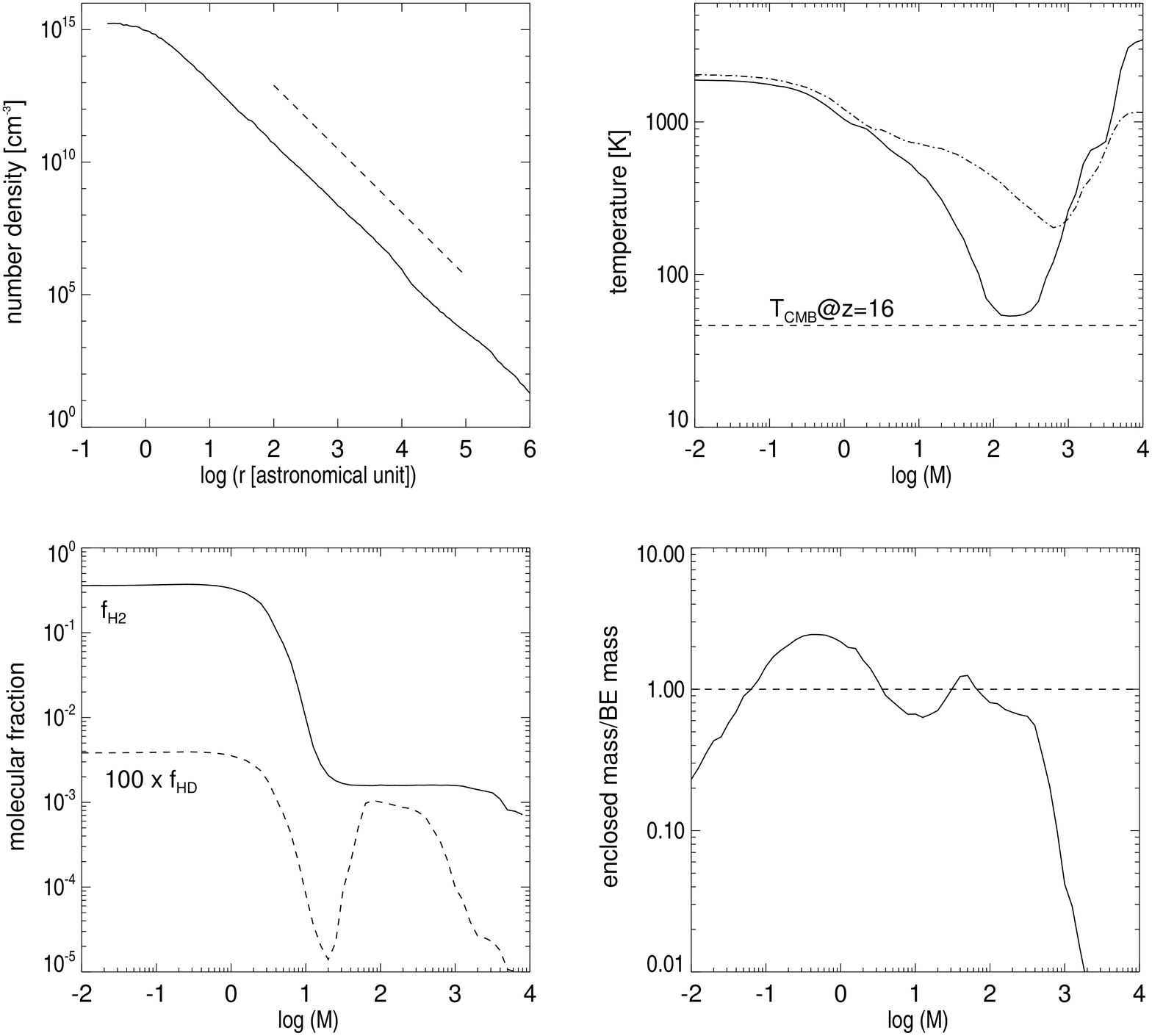}}
\caption{Radial profiles around the `second-generation star'
for density, temperature, 
molecular fraction, and the ratio of enclosed gas mass to
locally estimated Bonnor-Ebert mass.
The density is plotted as a function of distance from the center,
whereas the other three quantities are plotted as a function of 
enclosed gas mass. The dashed line in the density profile plot
indicates a power law of $\propto r^{-2.4}$. The horizontal
dashed line in the temperature profile plot indicates
the CMB temperature at $z=16$. We also show the temperature
profile of the primordial proto-star in Yoshida et al. (2006)
for comparison (dot-dashed line).
In the lower-left panel,
we shift the HD fraction two decades upward to simplify
comparison with \HH fraction.}
\label{fig:HDpanel}
\end{inlinefigure}

\begin{inlinefigure}
\resizebox{9cm}{!}{\includegraphics{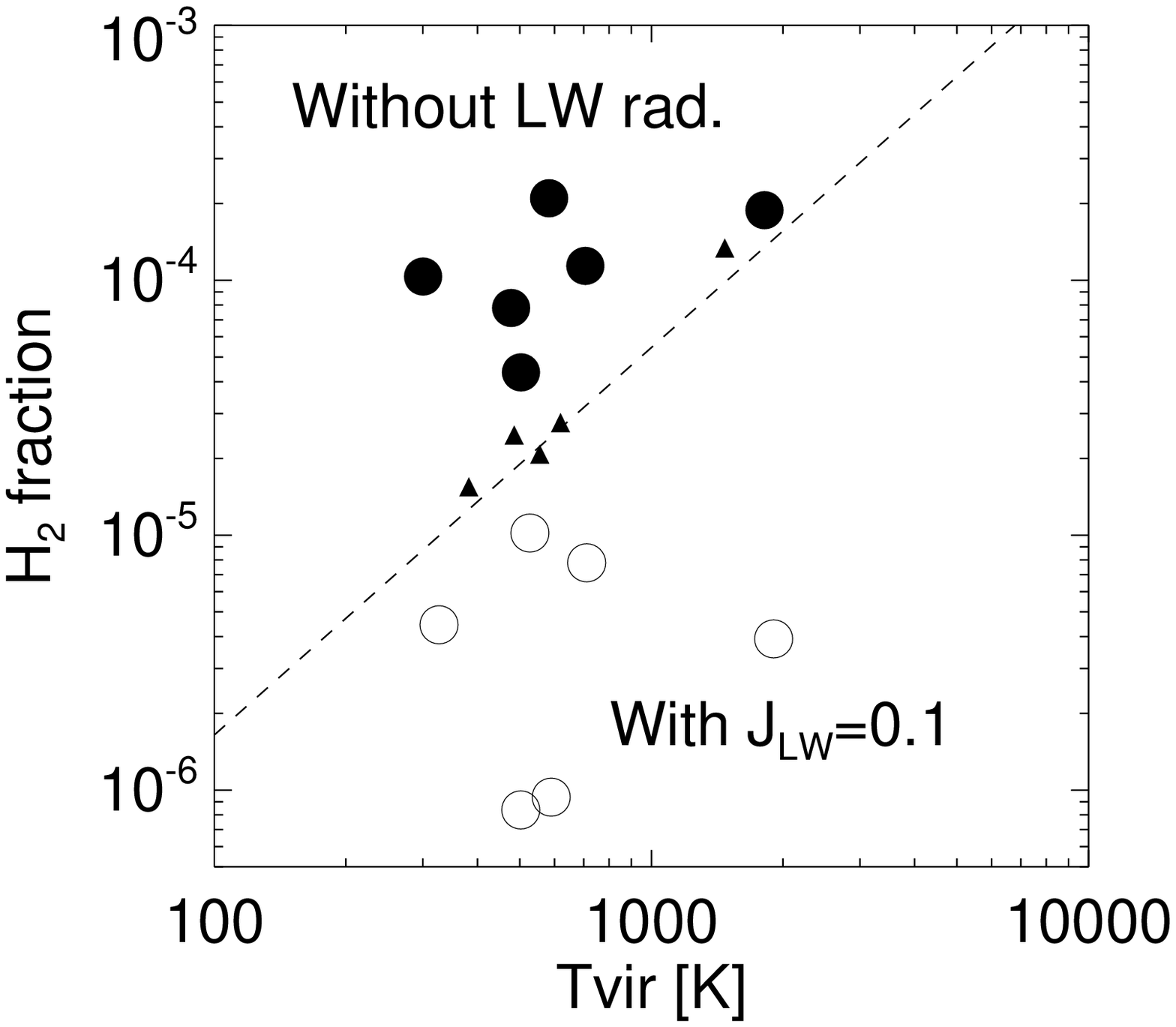}}
\resizebox{9cm}{!}{\includegraphics{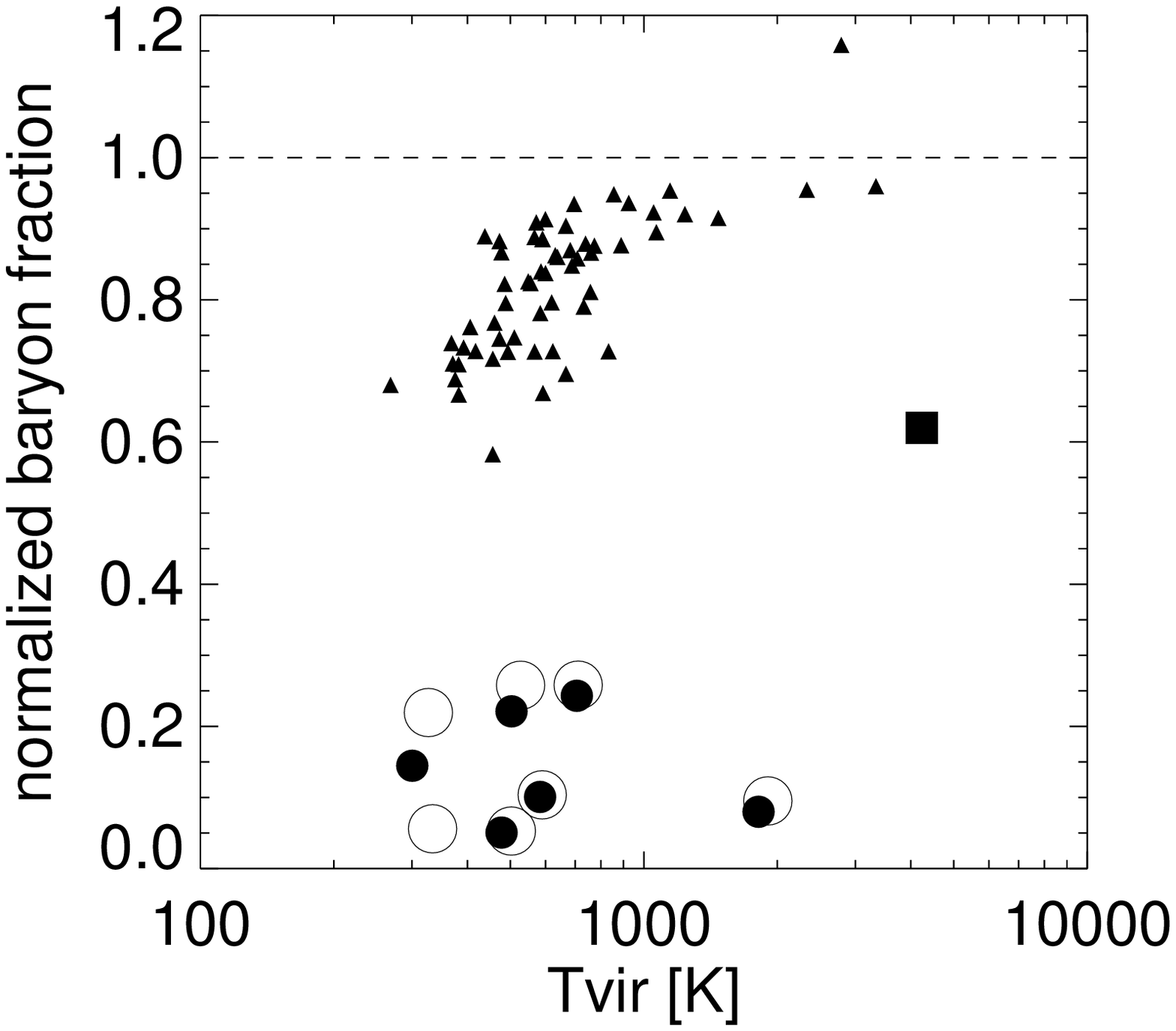}}
\caption{Mass-weighted mean H$_2$ fraction (left)
and the normalized baryon fraction (right) against virial temperature 
for halos in the relic \HII region at z=16. Filled circles are halos 
in the relic \HII region in the run without background radiation, 
whereas open circles are those in the run with $J_{\rm LW}$=0.1. 
We also show the molecular fractions and baryon fractions in the run 
without \HII region calculation (triangles). The dashed line in the top 
panel is the asymptotic \HH fraction in a collapsed neutral gas 
$\propto T^{1.52}$ derived by Tegmark et al. (1997).
The big square in the bottom panel is Halo C, which 
has a mixture of (once-)ionized gas and neutral gas.}
\label{fig:LWeffect}
\end{inlinefigure}

\subsection{The effect of ultra-violet radiation background}
\label{sec:LWbackground}
So far, we have not considered external radiation from other stars,
assuming that the first star in the simulation is the very first 
radiation source. We have also shown that recombination
radiation from the relic \HII region itself is unimportant.
In a more general cosmological context, there 
may be an early UV background radiation built up by stars/galaxies
that are formed elsewhere. In particular,
the intergalactic medium is nearly optically thin to 
photo-dissociating far-UV radiation\footnote{The primordial gas at very high 
redshift is not optically thin, because there is a trace amount of 
\HH molecules left over in the post-recombination epoch (e.g., Ricotti et al. 2001).
However, the residual \HH molecules in the diffuse intergalactic medium 
are quickly destroyed by the first luminous sources.}, 
and thus it is conceivable that an early far-UV background radiation
is quickly built up when stars/galaxies are formed. 
It is still unclear whether the net sign of feedback 
is positive (Ricotti, Gnedin \& Shull 2002) or 
negative (Oh \& Haiman 2003) in relic \HII regions 
relative to primordial neutral regions. The strength of the background LW 
radiation appears to play a key role in modulating between these regimes. 
Mesinger, Bryan \& Haiman (2006) recently examined these issues, 
and found that there is a critical value of 
$J_{\rm LW} \sim 0.01 \times 10^{-21} {\rm erg} {\rm s}^{-1} {\rm cm}^{-2} 
{\rm str}^{-1} {\rm Hz}^{-1}$ for which \HH cooling in relic 
\HII regions is strongly suppressed. However, their simulations do not 
take gas self-shielding into account, and thus the critical value they 
obtained is likely to be an under-estimate.

In principle, the background radiation itself evolves
and thus the intensity, as well as the spectrum, changes as
a function of time. Since the relevant energy range for
photo-dissociation is narrow, between 11.18 eV and 13.6 eV
\footnote{Photo-detachment of H$^-$ by lower energy photons
than LW photons could also affect the formation 
of \HH molecules, and hence it should, in principle, 
be taken into account. However, the relative intensities between the
relevant energy ranges are not well-constrained because of the uncertainties
in the sources of background radiation. We thus
ignore radiation below the LW bands as well, in order
to isolate the effect of photo-dissociation in the present paper.},
we assume that the spectrum is flat in this range, 
and we set the intensity to be 
$J(\nu) = 0.1 \times 10^{-21} {\rm erg} {\rm s}^{-1} {\rm cm}^{-2} 
{\rm str}^{-1} {\rm Hz}^{-1}$. 
We also implement the effects of \HH self-shielding,
as described in section \ref{sec:shield}. 

We evolve the system down to $z=16$ with the radiation on
and compare the outputs from those without background radiation.
Figure \ref{fig:LWeffect} shows the mean molecular fraction
in halos that are in the relic \HII region.
In the figure, open circles are for halos in the run with a LW radiation
field, 
and solid circles are for halos in the other run. 
For reference, we also show the molecular fractions in
the run without the \HII region calculation described in the 
previous section (triangles). 
The dashed line in the top panel in Figure \ref{fig:LWeffect} is 
the asymptotic molecular fraction
in a neutral primordial gas as calculated in Tegmark et al. (1997), 
which scales as $\propto T^{1.52}$. The molecular fraction
in the run without radiative transfer calculation is well 
described by the model.
The so-called positive feedback in relic \HII regions 
is clearly seen as enhanced large molecular fractions in the case
without LW radiation.
However, the positive feedback is effective 
{\it only if the background
far-UV radiation is weak}. Even though molecule formation
is promoted in relic \HII regions, UV background radiation 
with a modest intensity is enough to prevent the gas from
cooling and collapsing by molecular hydrogen cooling.
This is explained by the low-density, and equivalently
high entropy, of the gas in the relic \HII region.
Oh \& Haiman (2003) argue that photo-ionization heating
increases the gas entropy, which prevents gas condensation in 
low-mass halos, possibly acting as negative feedback.

\begin{inlinefigure}
\resizebox{17cm}{!}{\includegraphics{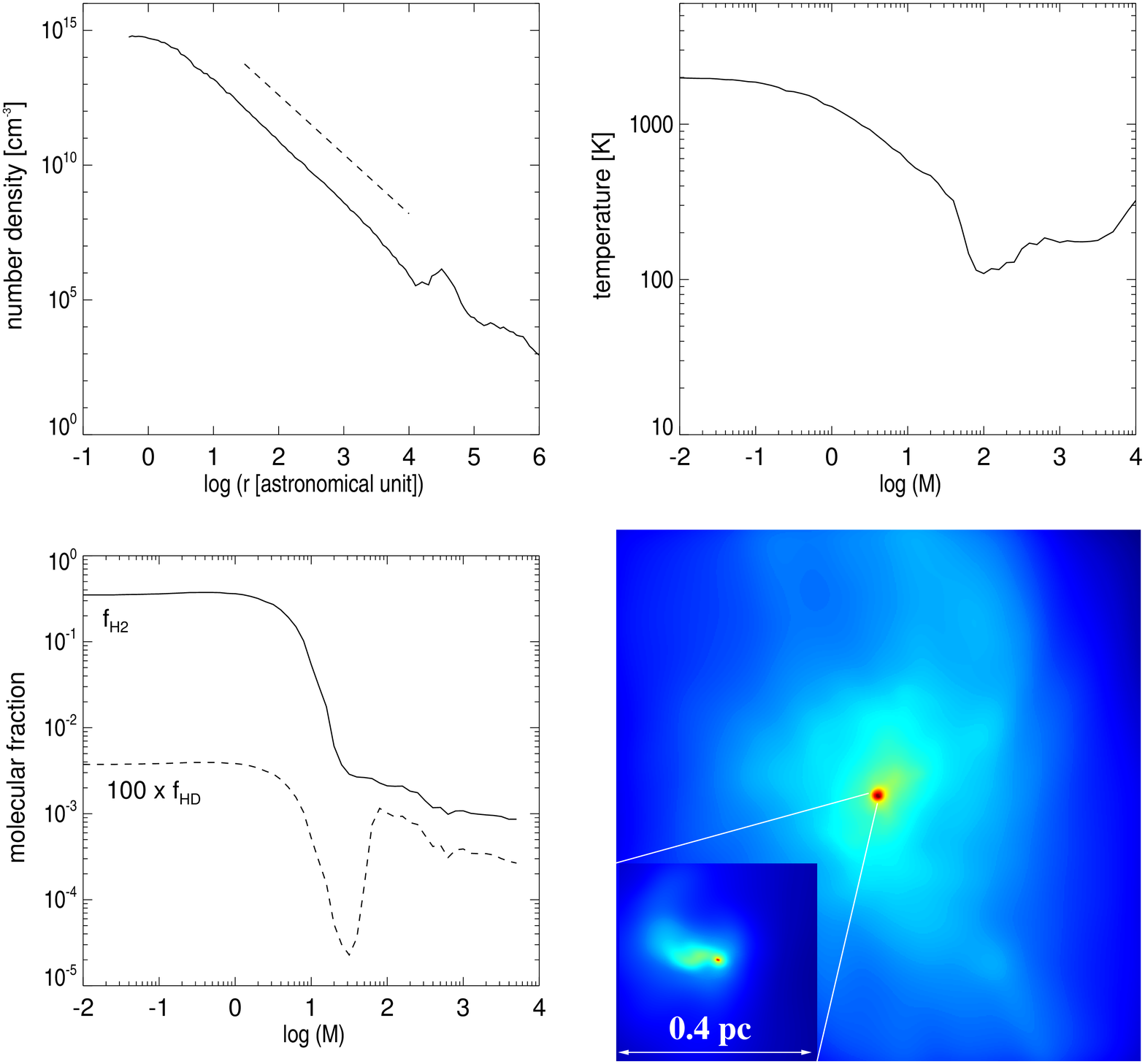}}
\caption{Radial profiles around the `second-generation star'
in the buried \HII region for density, temperature and 
molecular fraction. The lower-right panel shows the gas distribution
within the halo (the side length is $R_{\rm vir}=80$pc) and a close-up 
on the central 0.4 pc region. The dashed line in the density profile plot
indicates a power law of $\propto r^{-2.3}$.}
\label{fig:burried}
\end{inlinefigure}

The right panel of Figure \ref{fig:LWeffect} shows the normalized
baryon fraction for the `ionized' halos in the relic \HII region. 
Again, we show the baryon fractions for halos in the run
without the \HII region calculation (triangles).
Small-mass halos in the relic \HII region have a significantly
lower baryon fraction, and hence the mean gas density within them
is very small.
Consequently, molecule formation does not proceed
efficiently, and self-shielding does not counter-act photo-dissociation
for the adopted intensity of the background radiation.
Halos in the relic \HII region have a significantly
lower baryon fraction, and hence the mean gas density within them
is very small. Consequently, molecule formation does not proceed
efficiently, and self-shielding does not counter-act photo-dissociation.
Oh \& Haiman (2003) estimate that the gas fraction in mini-halos should
be as small as $\sim 0.1$, which is in good agreement
with the results of our three-dimensional simulations with hydrodynamics,
as is seen in Figure \ref{fig:LWeffect}.
The mean molecular fractions are below $10^{-5}$ in the case
with radiation, much less than the critical fraction necessary for gas 
cloud collapse (Yoshida et al. 2003; Oh \& Haiman 2003). It is interesting
that the fractions are also smaller than the asymptotic fraction (dashed line)
for a primordial gas, which indicates a net negative feedback effect in 
comparison with early primordial gas clouds.
We have set the radiation intensity to be a constant.
For a weaker UV radiation field, the result will be close
to the no radiation case, whereas for a stronger radiation
field,
photo-dissociation is more effective and thus the molecular
fraction will likely be smaller than shown in Fig. \ref{fig:LWeffect}. 
Note, however, that the formation and destruction processes are
highly non-linear, and hence the resulting molecular fraction
will not simply scale with the radiation intensity.

We close this section by mentioning that the effect of meta-galactic 
ionizing radiation
will be more profound in regulating star-formation in galaxies at high redshift.
A number of authors addressed this issue using one-dimensional
(Thoul \& Weinberg 1996; Kitayama \& Ikeuchi 2000)
and three-dimensional (Susa \& Umemura 2004a,b; 2006) 
calculations. These authors show that dwarf galaxy formation is regulated 
by both global UV background radiation and local UV sources. Once global 
reionization is completed, formation of \HH molecules is substantially suppressed
and thus it is likely that star-formation takes place only
in high-mass halos in which the gas cools by atomic processes.
As we have shown in this section, photo-dissociating radiation 
with a modest intensity is enough to delay star-formation in 
relic \HII regions {\it at least} for over tens of million years. 
If ionizing photons by global or local luminous UV sources 
arrive soon after the photo-dissociating photons, star-formation
in the region will be further delayed. 
%This is a complicated problem
%involving the epoch and topology of reionization within a
%very large cosmological volume, and hence we do not discuss the issue
%further in the present paper.

\subsection{Buried HII regions}
We have studied the formation and evolution of an \HII region
that extends much further than the host halo's virial radius.
There may be a distinct case that the I-front is trapped
well inside the host halo, making a compact \HII region.
Namely, if the I-front 
remains D-type during the stellar lifetime,
the size of the \HII region would be much smaller than the virial radius.
This corresponds to a case where the central star is less luminous, or
the host halo mass is very large.
KYSU simulated a case with a high mass halo, $M_{\rm halo} = 10^{7} M_{\odot}$,
and found that a D-type front travels
only up to $\sim 2$pc, and then falls back to the center.
It is interesting to study the evolution of such `buried' \HII regions
and the formation of second-generation stars in them.

To model this case, we use the same simulation as in the previous sections 
but switch off the star when the I-front reaches $\sim$ 5 pc
from the center. By this time, the central $\sim 4000 M_{\odot}$
part has been ionized. The ionized gas still has a high density
and thus recombines quickly, while expanding because of its thermal pressure, 
and the central part evolves essentially as in the isobaric 
case presented in section 2.
The mean gas density at the center is as large as $\sim 3\;{\rm cm}^{-3}$,
and the thermal energy is lost in less than a few million years 
mainly by hydrogen line cooling.
The gas re-collapses and condenses in about ten million years, 
and then a dense cold gas blob is formed at the center.
We again use the technique of Yoshida et al. (2006)
to follow the gas evolution to densities of $\sim 10^{15} {\rm cm}^{-3}$.
Figure \ref{fig:burried} shows the density, temperature and molecular fraction
profiles around the gas cloud. The lower-right panel shows the 
gas distribution within a volume of 100 pc on a side, and within 
the central 0.4 pc region. The extent of the compact \HII region 
can be seen as the central, roughly spherical part colored in red. 
The radial profiles look
similar to the second-generation object discussed in the previous section,
except that the minimum temperature is slightly higher, around $100$K.
This is because the collapsing cloud has a mixture of
an ionized gas and a neutral gas.
As soon as the central radiation source is switched off,
the sharp boundary (a pressure jump) of the buried \HII region 
drives significant mixing of the gas in both the ionized and neutral sides.
The two gases collapse toward the potential well of the host dark halo together,
and thus the final collapsing gas cloud contains a substantial amount
of neutral gas in and around the envelope. 
HD cooling does not operate in such a neutral gas packet,
maintaining 
its temperature just above 200K. When we calculate average temperatures 
over spherical shells, we obtain a slightly higher minimum temperature
because of the `warmer' neutral gas packets.
By calculating the ratio of the enclosed gas mass to the locally
estimated Bonnor-Ebert mass (equation [\ref{eq:BEmass}]),
we find that the mass of the cloud that is run-away collapsing is $\sim 80M_{\odot}$,
roughly consistent with the cloud mass in the re-collapse calculation in section
\ref{sec:secondstar}.
Throughout the evolution, we observe no sign of fragmentation of the cloud 
up to the final output time when the cloud core density reaches $\sim 10^{15} {\rm cm}^{-3}$. 
Following the argument of Omukai \& Palla (2003) and Yoshida et al. (2006), 
we expect the central proto-star in this buried \HII region will grow, 
to be a massive metal-free star.

\section{Summary and Discussion}
\label{sec:summary}

We have studied the formation and structure of early cosmological
\HII regions using three-dimensional radiation hydrodynamics 
simulations. We have also examined in detail the formation of 
second-generation stars, by following cooling and condensation 
of primordial gas in relic \HII regions. With the three-dimensional 
simulations, we can eliminate a number of uncertain assumptions 
regarding the formation of early \HII regions. First, the simulations 
allow us to directly calculate the ionizing photon escape fraction from 
cosmological minihalos. With enough mass resolution, our calculation 
self-consistently includes the effect of gas clumping.
We obtain photon escape fractions of order unity, for the cases with 
rather massive ($M_* > 60 M_{\odot}$) central stars.
While our radiation tranfer scheme is based on a few approximations,
the accuracy of the method is tested against the results of 
previous one-dimensional calculations (KYSU; Whalen et al. 2004).
We conclude that the technique is well-suited for the particular
problem of early \HII regions.

%revised by naoki
We find that a large volume of a few kiloparsec 
diameter is ionized by a single massive star, 
within which a smaller \HeIII region is embedded. 
The ionized hot gas is evacuated at a velocity of $\sim 30$ km/sec.
After the central star dies off, the gas recombines, and cools first by 
atomic line cooling, then by \HH cooling, and finally by HD line cooling 
down to a few tens Kelvin. The fractionation of HD/\HH occurs when the 
gas temperature becomes as low as 100 K, and the HD fraction reaches $10^{-5}$.
Run-away collapse is triggered when the cloud mass exceeds
$\sim 40 M_{\odot}$ at the temperature minimum of $\sim 40-50$K.
It is substantially smaller than the corresponding mass scale 
for the first generation stars, $\sim 300 M_{\odot}$ (Yoshida et al. 2006), 
indicating that second-generation stars formed under these conditions 
likely have smaller masses than the first stars. At least
the maximum mass will be limited by the gas clump mass.
It is worth noting that the elemental abundance patterns
of hyper metal-poor stars 
recently discovered by Christlieb et al. (2002) and Frebel et al. (2005) 
indicate that the early metal-enrichment was caused by supernova explosions
with a progenitor mass of $\sim 25 M_{\odot}$ (Umeda \& Nomoto 2003; Iwamoto et al. 2005).
We speculate that the second-generation stars formed in relic \HII
regions likely have a similar mass.
%revision end 

It is important to note that the re-incorporation and recollapse of the 
ionized gas happens very late. Full re-incorporation of the gas is not achieved
even after a hundred million years, but a small amount of the gas at 
the center of a growing halo recollapses to form a star-forming 
gas cloud in a hundred million years. 
O'Shea et al. (2005) report a shorter time scale for recollapse.
The discrepancy can be understood from the fact that, in their calculation, 
the density structure is unchanged 
during the evolution of the \HII region. The highest density regions have 
the shortest recombination times and radiative cooling times, and thus the gas evolves 
faster than it should if hydrodynamics and radiation are coupled.
By employing a more accurate treatment of radiation transport, 
we are able to obtain a realistic estimate for the
timescale of suppression of star-formation in early halos. 
We also examined the feedback effect of external UV radiation, 
and found that it strongly reduces the \HH fraction and baryon 
fraction in low-mass halos.  
Our results suggest that the long-term evolution and the fate of
the ionized gas in and outside relic \HII regions are determined
by a rather complex combination of many factors such as the source
luminosity, relative positions, exact timing of collapse, 
strength of background radiation etc., and hence it is very difficult to 
establish a `rule' of the sign of the net feedback effect. 
This is also supported by recent numerical studies by Ahn \& Shapiro (2007) 
and Susa \& Umemura (2006) who explored a large parameter space
in order to examine the net effect of photo-dissociation.
We argue that it is necessary to carry out a large cosmological simulation 
with all these relevant physics implemented, as done in the present paper, 
in order to to follow the evolution in relic \HII regions; i.e., 
probable sites of (proto-)galaxy formation.

%newly added by naoki
In the present paper, we did not consider cooling by heavy elements,
assuming that the gas in \HII regions remains chemically pristine
although its ionization fraction changes.  Metal-enrichment can be 
caused not only by supernova explosions but also by stellar winds 
during the evolution of the central massive star via mixing and 
self-enrichment (e.g., Vink \& de Koter 2005). 
It is thus conceivable that at least a small amount of carbon
can be dispersed in the central part of the \HII region. It is well 
known that trace amounts of heavy elements such as carbon and oxygen 
can significantly change the gas cooling efficiency (e.g., 
Bromm \& Loeb 2004; Omukai et al. 2005;  Santoro \& Shull 2006). 
Intriguingly, however, Omukai et al.'s calculations show that
the thermal evolution of a collapsing gas with a modest metal content, 
but without dust, is similar to that found in our simulation of an 
ionized primordial gas with HD cooling. Although radiative cooling
by the C{\sc ii} fine structure transition can bring the gas temperature 
as low as 10 K, whereas HD cooling is efficient at $T>30$ K, 
the minimum gas temperature is limited by $T_{\rm CMB} \ga 30$ K 
at $z>10$, and thus gas-phase metal enrichment may not substantially
affect the results of our calculations. 
We argue that the existence of dust may play a more important role 
in significantly changing the characteristic mass of collapsing
gas clouds. The gas temperature can be kept low at high 
densities by dust thermal emission, which brings the collapse
mass scale to the `opacity-limited' value of $< 1 M_{\odot}$
(Low \& Lynden-Bell 1976; Schneider et al. 2006).
%end

We now discuss the implications of our results for early galaxy formation.
While the very first generation stars are likely to be 
born in low-mass dark halos, it is often argued that 
efficient star formation takes place only in large halos 
in which the gas cools via atomic processes  (e.g., Barkana \& Loeb 2000).
A crude assumption in such analyses is that the gas distribution
traces that of dark matter; i.e., the baryon fraction is close
to the cosmic mean, so that the gas cools
efficiently at the centers of dark halos. 
Our simulation results invalidate these assumptions. 
If the first stars are massive (not necessarily very massive), 
they decouple the gas distribution from that of dark matter for as 
long as a Hubble time at $z\sim 20$. 
We argue that efficient star-formation does not take place
immediately in dwarf (proto-)galaxies
whose progenitors have hosted massive PopIII stars earlier.

The first stars may leave remnant black holes
at the end of their lives, unless they trigger 
complete disruption by pair-instability supernova explosions. 
Efficient accretion onto the remnant black holes
could lead to formation of miniquasars (Kuhlen \& Madau 2005).
Our simulation results indicate, however, that accretion 
is rather inefficient at least for some tens of millions years
because the progenitors, massive stars,
emit a large number of UV photons and the surrounding gas
is evacuated out of the gravitational potential well of 
the host halo. It will be interesting to simulate the 
inefficient accretion process onto the remnant Population III 
black holes.

The first \HII regions may imprint a distinct signature
as a source of CMB secondary anisotropies via the kinematic
Sunyaev-Zel'dovich effect. 
The diameter of a few kiloparsec proper region at $z=10-20$
is about a few arcsecs in angular size. Hence,
if a large number of early \HII regions are formed,
they will produce CMB temperature fluctuations, 
at arcsec scales in the angular power spectrum. 
If the first \HII regions are clustered in cosmological 
high-density peaks such as those studied by Reed et al. (2005),
they will make a large \HII bubble with a mega-parsec diameter 
at $z>10$, which may possibly be detected by ALMA
\footnote{http://www.nro.nao.ac.jp/alma/} with multiple channels 
with a very long-time exposure.
Finally, it has been long known that neutral hydrogen at high redshift 
may be directly detectable through the redshifted 21cm line
(see Furlanetto, Oh \& Briggs (2006) for a recent review).
This can be seen either in absorption or emission against the CMB.
We have made a brightness temperature map following Nusser (2005) 
using the simulation outputs. 
We find that the overall signal is a strong function of time (equivalently
the evolutionary phase of the \HII region).
While the \HII region during the central stellar lifetime 
appears essentially as a ``void'' in the map, because the HI content is
extremely small, the relic \HII region
can be seen as a strong emission source with the differential 
temperature against the CMB up to $\delta T_{b} \sim +50$mK
for an angular resolution of $\sim 0.02"$. While it is unlikely 
that currently planned radio telescopes can directly detect individual
\HII patches, the fluctuations in the 21cm background from early \HII regions
can be significantly enhanced just like those from cosmological minihalos
(e.g., Shapiro et al. 2006). We will study the observability
of early \HII regions in future work.

\bigskip
\bigskip

NY thanks Volker Bromm, Simon Glover, Tom Abel, and Dan Whalen 
for helpful discussions on chemistry and radiative transfer.
The authors are grateful to Hajime Susa and Kaz Omukai for a careful
reading of the draft of this paper. 
We also thank the anonymous 
referee for giving many constructive comments. 
The simulations were performed at the Center
for Parallel Astrophysical Computing at Harvard-Smithsonian
Center for Astrophysics, at the Center
for Computational Cosmology at Nagoya University,
and at the Data-Reservoir at the University of Tokyo.
NY thanks Mary Inaba and Kei Hiraki at the University
of Tokyo for providing the computational resources.
The work is supported in part by the 21st Century
ORIUM Program at Nagoya University, by The Mitsubishi
Foundation, and by the Grants-in-Aid 
for Young Scientists 17684008 (NY) and 18740112 (TK). 
by the Ministry of Education, 
Culture, Science and Technology of Japan (NY).
SPO gratefully acknowledges NSF grant AST-0407084 and NASA grant 
NNG06GH95G for support.

\clearpage
\appendix
Chemical reactions involving H$_3^+$ and HeH$^+$
that may be important in a primordial gas are summarized below.

For H$_3^+$, the main formation paths are 
\begin{equation}
\Htwoplus + \Htwo \rightarrow \Hthreeplus  + {\rm H},
\end{equation}
and 
 \begin{equation}
\Htwoplus + {\rm H} \rightarrow \Hthreeplus  + \gamma.
\end{equation}
We include the reverse reaction of these
and dissociative recombination
\begin{eqnarray}
\Hthreeplus + e &\rightarrow& \Htwo  + {\rm H}, \\
&\rightarrow& 3{\rm H}.
\end{eqnarray}

For HeH$^+$ molecules, the main formation paths are the radiative
association
\begin{equation}
{\rm He} + \Htwo \rightarrow \HeHII + {\rm H}^+ ,    %radiative association
\end{equation}
and the inverse rotational predissociation
\begin{equation}
{\rm He}   + {\rm H}^+  \rightarrow \HeHII + \gamma.     % inverse rotational predissociation
\end{equation}
We include the reverse processes and the dissociative recombination
\begin{equation}
\HeHII   + e  \rightarrow {\rm He} + {\rm H},   
\end{equation}
and a formation path of H$_3^+$
\begin{equation}
\HeHII + \Htwo  \rightarrow \Hthreeplus  + {\rm He}
\end{equation}
The reaction rates are taken from Stancil et al. (1998).
Hirata \& Padmanabhan (2006) recently revised post-recombination
calculations of the early Universe. They argue that H$_2^+$ formation
via
\begin{equation}
\HeHII + {\rm H} \rightarrow \Htwoplus  + {\rm He}
\end{equation}
is the dominant path at high redshifts ($z>200$). 
We have checked that this process does not affect 
the  H$_2^+$ abundance in all the thermal phases and evolution
considered in this paper.
Also, photo-dissociation by the cosmic microwave background radiation 
is unimportant in the redshift range we consider.

We have run the same calculations of an isobarically cooling gas
as in section \ref{sec:isobaric} including the above species and reactions.
We conclude that the number fractions of these ionic molecules are
extremely small, and do not affect the thermal evolution 
of the gas. Cooling by H$_3^+$ ions can potentially be important
in collapsing gas clouds at very high densities, 
$n_{\rm H} \sim 10^{7} - 10^{10} {\rm cm}^{-3}$ when there is
some (weak) ionization source (S. Glover, private communication),
but it is outside the regime and conditions we consider 
in the present paper.

 \begin{deluxetable}{llll}
 %\scriptsize
 %\rotate

 \tablecolumns{4}
 \tablecaption{REACTION RATE COEFFICIENTS: Deuterium Chemistry}
 \tablehead{
 &\colhead{Reactions} & \colhead{Rate Coefficients
  (cm$^3$s$^{-1}$)} 
 & \colhead{Reference}}
 \startdata
   % D recombination
   (D1) & ${\rm D^+}+ e \rightarrow {\rm D}+h\nu$
 & $k_{\rm D1} = 3.6 \times 10^{-12} (T/300)^{-0.75}$ & 1 \\
 \\
   % D H+ charge transfer
   (D2) & ${\rm D}+ {\rm H^+} \rightarrow {\rm D^+}+{\rm H}$
 & $k_{\rm D2} = 2.0 \times10^{-10} T^{0.402} \exp (-37.1/T) - 3.31\times 10^{-17} T^{1.48} $  & 2 \\
 \\
   % D+ H charge transfer
   (D3) & ${\rm D^+}+ {\rm H} \rightarrow {\rm D}+{\rm H^+}$
 & $k_{\rm D3} = 2.06 \times10^{-10} T^{0.396} \exp (-33.0/T) + 2.03\times 10^{-9} T^{-0.332} $  & 2 \\
 \\
   (D4) & ${\rm D}+ {\rm H} \rightarrow {\rm HD}+h\nu$
 & $k_{\rm D4} = 1.0 \times10^{-25}$ & 3 \\
 \\
   (D5) & ${\rm D}+ {\rm H_2} \rightarrow {\rm H}+{\rm HD}$
 & $k_{\rm D5} = 9.0 \times 10^{-11} \exp \left(-3876/T\right)$
 & 4 \\
 \\
   (D6) & ${\rm HD^+}+ {\rm H} \rightarrow {\rm H^+}+{\rm HD}$
 & $k_{\rm D6} = 6.4 \times 10^{-10}$ & 3 \\
 \\
   (D7) & ${\rm D^+}+ {\rm H_2} \rightarrow {\rm H^+}+{\rm HD}$
 & $k_{\rm D7} = 1.6 \times 10^{-9}$ & 4 \\
 \\
   (D8) & ${\rm HD}+ {\rm H} \rightarrow {\rm H_2}+{\rm D}$
 & $k_{\rm D8} = 3.2 \times 10^{-11} \exp \left(-3624/T\right)$ & 3 \\
 \\
   (D9) & ${\rm HD}+ {\rm H^+} \rightarrow {\rm H_2}+{\rm D^+}$
 & $k_{\rm D9} = 1.0 \times 10^{-9} \exp \left(-464/T\right)$ & 3 \\
 \\
   (40) & ${\rm D}+ {\rm H^+} \rightarrow {\rm HD^+}+h\nu$
 & $k_{\rm D10} = {\rm dex} [-19.38 -1.523\log T+1.118(\log T)^2 -0.1269(\log T)^3]$ & 3 \\
 \\
   (D11) & ${\rm D^+}+ {\rm H} \rightarrow {\rm HD^+}+h\nu$
 & $k_{\rm D11} = {\rm dex} [-19.38 -1.523\log T+ 1.118(\log T)^2 -0.1269(\log T)^3]$ & 3 \\
 \\
   (D12) & ${\rm HD^+}+ {\rm e} \rightarrow {\rm H}+{\rm D}$
 & $k_{\rm D12}=7.2 \times 10^{-8}T^{-1/2}$ & 3 \\
 \\
   (D13) & ${\rm D}+ e \rightarrow {\rm D^-}+h\nu$
 & $k_{\rm D13}=3.0 \times 10^{-16}(T/300)^{0.95}\exp(-T/9320)$ & 1 \\
 \\
   (D14) & ${\rm D^+}+ {\rm D^-} \rightarrow 2{\rm D}$
 & $k_{\rm D14}=5.7 \times 10^{-8}(T/300)^{-0.5}$ & 1 \\
 \\
   (D15) & ${\rm H^+}+ {\rm D^-} \rightarrow {\rm D}+{\rm H}$
 & $k_{\rm D15}=4.6 \times 10^{-8}(T/300)^{-0.5}$ & 1 \\
 \\
   (D16) & ${\rm H^-}+ {\rm D} \rightarrow {\rm H}+{\rm D^-}$
 & $k_{\rm D16}=6.4 \times 10^{-9}(T/300)^{0.41}$ & 1 \\
 \\
   (D17) & ${\rm D^-}+ {\rm H} \rightarrow {\rm D}+{\rm H^-}$
 & $k_{\rm D17}=6.4 \times 10^{-9}(T/300)^{0.41}$ & 1 \\
 \\
   (D18) & ${\rm D^-}+ {\rm H} \rightarrow {\rm HD}+e$
 & $k_{\rm D18}=1.5 \times 10^{-9}(T/300)^{-0.1}$ & 1
 \enddata
 \tablecomments{
 (1): Galli \& Palla (1998);
 (2): Savin (2002);
 (3): Stancil, Lepp \& Dalgarno (1998)
 (4): Wang \& Stancil (2002); 
 }
 \label{tab:hd rate}
 \end{deluxetable}

\end{document}